\documentclass{article}

\usepackage[english]{babel}

\usepackage[numbers]{natbib}

\usepackage[letterpaper,top=2cm,bottom=2cm,left=3cm,right=3cm,marginparwidth=1.75cm]{geometry}

\usepackage{amsmath}
\usepackage{amssymb}
\usepackage{mathrsfs}
\usepackage{graphicx}
\usepackage[colorlinks=true, allcolors=blue]{hyperref}
\usepackage{authblk}

\usepackage{wrapfig}
\usepackage{bm} 
\usepackage{colortbl} 
\usepackage[table]{xcolor}
\usepackage{makecell} 
\usepackage{rotating}
\usepackage{multirow}

\usepackage{placeins}

\usepackage{subcaption}

\title{{\fontfamily{lmtt}\selectfont DeepCormack} : Fermi surface tomography using model-based data-driven algorithms}

\author{\normalsize{Georg F. B. Lovric{\footnotesize$^{1}$}, Bryn Drury{\footnotesize$^{2}$}, Carola-Bibiane Schönlieb{\footnotesize$^{1}$}, Stephen B. Dugdale{\footnotesize$^{2}$}, Ander Biguri{\footnotesize$^{1}$}} \\
\footnotesize{\textit{$^1$Department of Applied Mathematics and Theoretical Physics, University of Cambridge, Cambridge, UK.} \\ \vspace{0.05cm} \textit{$^2$H.H. Wills Physics Laboratory, University of Bristol, Tyndall Avenue, Bristol BS8 1TL, UK}}}

\begin{document}
\maketitle

\begin{abstract} 
The experimental reconstruction of the 3D two-photon momentum density (TPMD) via angular correlation of electron-positron annihilation radiation (ACAR) is a particularly useful method for studying material Fermi surfaces. It does not rely on low temperatures, UHV conditions, or strong magnetic fields, and enables the study of the spin-resolved electronic structure of materials. Yet, it remains a challenging inverse problem. Typically, $\sim10^{8}$ positron annihilation events are measured for $3-6$ projections of the TPMD at different angles. The standard reconstruction approach is an ACAR adaptation of Cormack's method (the MCM) that leverages the inherent symmetry in the crystal's structure. However, the poor signal-to-noise ratio means collecting data of sufficient quality for Fermi surface studies can take months per sample. Even at high counts, instrumental smearing typically requires additional deconvolution methods to sharpen the density and uncover Fermi surface features.

We present {\fontfamily{lmtt}\selectfont DeepCormack}, a family of data-driven model-based reconstruction algorithms that augments the MCM by integrating supervised deep-learning models (CNN, MLP, and UNet) at various stages. To overcome the lack of large experimental training sets, we propose a method which leverages singular value decomposition with dynamic mode decomposition to generate realistic synthetic TPMD volumes, requiring only a single reference momentum density computed via density functional theory (DFT) \cite{Ernsting_2014}.

On synthetic test data, {\fontfamily{lmtt}\selectfont DeepCormack} improves reconstruction quality over MCM by about $8.5$ dB PSNR at standard counts (200M) and remains stable at reduced counts: PSNR decreases from $40.69 \pm 3.61$ dB at $200$M to $38.22 \pm 3.60$ dB at $10$M, compared with $32.38 \pm 4.14$ dB to $26.12 \pm 4.38$ dB for MCM across the same range, enabling significantly faster acquisition times. On experimental data, generalisation depends strongly on how well the training distribution matches the sample. For the copper reference momentum density under simulated experimental conditions at $200$M counts the best model reaches $40.05 \pm 3.75$ dB PSNR versus $31.62 \pm 3.00$ dB for MCM. However, model performance degraded significantly when testing on an out-of-distribution sample of $\text{ZrZn}_2$. We therefore recommend pairing {\fontfamily{lmtt}\selectfont DeepCormack} with a DFT calculation of the target material to create sample-specific training data. Overall, our proposed workflow and method offers either much higher quality reconstructions, or enables significantly faster ones, in the order of weeks.
\end{abstract}

\section{Introduction \label{Introduction}}
The Fermi surface of a material is the surface in reciprocal space (or \textbf{k}-space) separating occupied from unoccupied electronic states at zero temperature ($T=0 {\rm K}$). Its shape is determined entirely by the electronic band structure, i.e. the relationship between electron energy and crystal momentum evaluated at the chemical potential. This shape governs many of the key properties of a metal, including electrical and thermal conductivity, magnetism, and more exotic phenomena such as superconductivity. The Fermi surface is not physical, it is a representation of which electronic states are available for low-energy excitations. Almost all low-energy physics in a metal is controlled by electrons near the Fermi surface: states deeper within it are frozen by the Pauli exclusion principle and cannot participate in transport, magnetic ordering, or Cooper pairing \cite{kittel, Dugdale_2016}. In the free-electron model, the Fermi surface is a simple sphere in \textbf{k}-space, but in real materials it becomes more complex due to the periodic crystal potential leading to electronic band structures that disperse anisotropically. This can lead to Fermi surfaces which comprise multiple interconnected sheets with topologies which are consistent with the underlying crystal symmetry \cite{Dugdale_2016}. These deviations in the Fermi surface topology, in comparison to the free-electron picture, can play a pivotal role in the physical properties of metals, for example, influencing the appearance of superconductivity, magnetic order and charge-density waves \cite{Dugdale_2016}. The geometry of the Fermi surface therefore encodes a great deal about what a metal can do, and having more accurate methods for measuring it is therefore important. 

Angular correlation of annihilation radiation (ACAR) is one such method \cite{Dugdale_2014,KontrymSznajd_2009}. It is also a method which is bulk sensitive, can be applied at room temperature, and which is suitable for disordered alloys \cite{Dugdale_2024,PhysRevB.69.174406}. When positrons are implanted into a sample, they thermalise quickly before annihilating with an electron, emitting two photons. As the thermal positrons carry negligible momentum at the point of annihilation, the small angular deviations of the photon pair from exact collinearity (on the order of milliradians) essentially encodes the momentum of the electron immediately prior to annihilation. In a 2D-ACAR experiment, two position-sensitive detectors placed on either side of the sample record the coordinates of the coincident gamma-photons which impinge on each detector, allowing the angular deviations of each detected annihilation event to be recorded. Accumulating on the order of $\sim 10^8$ coincidence events produces a 2D projection of the underlying three-dimensional two-photon momentum density (TPMD). By repeating this at different sample orientations (corresponding to different crystallographic directions), a set of projections is obtained from which the full 3D TPMD can be reconstructed. The features of the Fermi surface geometry can be extracted from the reconstructed density by converting from \textbf{p}-space to \textbf{k}-space via the Lock-Crisp-West (LCW) theorem \cite{LCW_Theorem, Dugdale_2014}. This LCW procedure sums over reciprocal lattice vector contributions to reduce the extended-zone momentum density into the reduced Brillouin zone, giving us the \textbf{k}-space density which represents the occupation number {\it as seen by the positron}. The Fermi surface then appears as a discontinuity in this occupation number, and can be recovered by, for example, computing the gradient. Some further details about the experimental and theoretical background to the measurement of the Fermi surface by the 2D-ACAR method can be found in Refs.~\cite{2D_ACAR_PhD_Thesis, Acquisition_Time_Dugdale2013}.

Recovering the 3D TPMD from a small number of projections remains a challenging inverse problem. Several reconstruction algorithms have been developed for this purpose, including filtered backprojection, spherical harmonic methods \cite{pecora_1989, Pecora_SphHarm}, and more recently regularisation-based approaches \cite{weber_2014, Weber_2015, Leitner2016}. In this work, we focus on the planar reconstruction method originally proposed by Cormack \cite{cormack_1963, cormack_1964, AMCormack1973} and later adapted for Fermi surface studies by Kontrym-Sznajd \cite{modified_cormack_method}. The \textit{modified Cormack method} (hereafter MCM) is well-suited to 2D-ACAR data because its least-squares properties handle experimental errors through natural truncation of the series, and because it enforces a consistency condition on the projection data that can correct for inconsistencies introduced by noise or normalisation errors. The underlying method uses polar Fourier series, expanding the projection data into sets of orthogonal polynomials (Chebyshev and Zernike), to carry out the reconstruction. The inherent symmetry of the crystals being studied and the relatively smooth nature of the reconstructed densities, make it a particularly powerful technique for recovering information about the Fermi surface.

Despite the advantages of the MCM, acquiring 2D-ACAR data of sufficient quality for Fermi surface studies is extremely time-consuming in practice. To make measurements with sufficiently high angular resolution (recalling that it is the angle that is proportional to the electron momentum), the joint requirements of high detector efficiency and acceptable position resolution mean that there can be up to 24m between the two detectors. The solid angle subtended by the detectors is therefore very small, resulting in relatively low coincidence count rates in the range of hundreds of counts per second. There is an inherent trade-off between acquisition time and that overall statistical precision of a spectrum, with the collection of more annihilation events reducing noise but requiring longer measurements. With a radioisotope source (typically $^{22}$Na), obtaining a single high-quality data set comprising five to six projections at $200$ million counts each can takes approximately three to four months. A further complication arises for reconstruction algorithms which operate in a slice-by-slice manner: as the number of recorded counts decreases away from the centre of the projections (corresponding to zero momentum), noise will vary systematically across the reconstructed slices. This results in the reconstructed densities at points away from the centre of the momentum density having much lower quality than ones closer to it. An approach which can simultaneously improve the momentum density reconstruction quality from ACAR data while at the same time lowering the acquisition time (meaning noisier projections) would therefore be desirable.

Given that 2D-ACAR shares its core measurement geometry with positron emission tomography (PET), there is good reason to expect that machine learning advances in medical image reconstruction could translate productively to Fermi surface tomography, given the significant advances in the field in the last few years, e.g. \cite{DeepFBP,10729663,Arridge,Maximilian}. The goal of this work is to investigate whether similar approaches can positively translate to ACAR tomography, using supervised neural networks in conjunction with classical reconstruction techniques to improve the recovered Fermi surface details.

To this end, we introduce {\fontfamily{lmtt}\selectfont DeepCormack}, a hybrid framework combining the Cormack-Kontrym-Sznajd method with supervised deep learning. The network includes a 1D convolutional neural network trained in projection space, a multi-layer perceptron in radial density function space, and a 2D UNet in image space, learning corrections that reduce noise and convolution artefacts while partially compensating for the information loss inherent in the limited angular sampling from $5$ projections. As reconstruction quality degrades away from the centre of a momentum density, the effect of conditioning the UNet on the statistical precision (characterised by the total number of counts in a spectrum) to make it robust to the variation in noise is also studied.

To support training and evaluation, we also propose a data-driven simulation framework based on singular value decomposition (SVD) and dynamic mode decomposition (DMD), which enables the generation of realistic, continuously varying synthetic 3D TPMD volumes, based entirely on a DFT-generated reference momentum density for copper \cite{Ernsting_2014}. Copper was chosen for its simple but anisotropic Fermi surface topology, as well as having the same crystal structure (face-centred cubic) as the intermetallic compound ZrZn$_2$, for which we have DFT calculations of the TPMD available. This represents an excellent intermediate step between the relatively simple momentum density of copper (with its single Fermi surface sheet) and the challenge of reconstructing from real high quality experimental data (which have been measured by Major {\it et al.} \cite{Major_2004}). With a much larger lattice constant (factor of two) compared to copper, the impact of resolution on the features across the much smaller Brillouin zone is likely to be more significant. Furthermore, with multiple Fermi surface sheets, its momentum density is quite different from copper which provides an excellent test of the training data based on copper. Together, these components form a physically consistent and scalable pipeline for machine learning-assisted ACAR reconstruction under experimentally realistic conditions. 

By investigating the application of modern deep learning tools in ACAR tomography, our hope is to open up the area of Fermiology to the wider inverse problem community. Finding techniques that allow for improved reconstruction while lowering the required statistical precision of the projections can enable faster and more efficient characterization of a material's properties, and make the study of Fermi surfaces more accessible.

The paper is organised as follows: Section \ref{Methods} presents the main algorithm and data generation approach. Section \ref{Results} presents experimental results of {\fontfamily{lmtt}\selectfont DeepCormack} for different model configurations, across acquisition counts ($10 - 200$M), and for an out-of-distribution (OOD) sample of ZrZn$_2$ to test its ability to generalize. Section \ref{Discussion} discusses these results in depth, while Section~\ref{Conclusions} summarises the main findings, discusses current limitations, and outlines improvements to the method. 

\paragraph{The main contributions can be summarized as follows: \label{Main_Contributions}} 

\begin{itemize}
 \item A derivation of the modified Cormack method reconstruction method in operator notation for the field of inverse problems.
 \item Novel method using SVD and DMD for generating sufficiently realistic 3D-TPMD data to train ML models with;
 \item Framework for integrating neural networks into the modified Cormack method for substantially improved Fermi surface reconstruction with ACAR tomography, sufficiently flexible enough to allow for slice-by-slice reconstruction despite varying noise levels across the TPMD volume; 
 \item Practical demonstrations of {\fontfamily{lmtt}\selectfont DeepCormack}'s utility, on both pseudo-real data and DFT-generated TPMD data under simulated experimental conditions.
\end{itemize}

\section{Methods \label{Methods}}
We propose a family of methods which integrate neural networks into the modified Cormack method (described in section \ref{Methods: The MCM Math}) to improve the two-photon momentum density reconstruction that can be achieved via conventional angular correlation of electron-positron annihilation radiation measurements. This framework, which we have dubbed {\fontfamily{lmtt}\selectfont DeepCormack} (\ref{Methods: DeepCormack Design}), requires supervised learning. Section \ref{Methods: Data Simulation} proposes a data simulation method which allows for paired ground truth and simulated measurement (synthetic) 3D TPMDs to be generated at scale.

\FloatBarrier
\subsection{Fermi Surface Tomography \label{Methods: The MCM Math}}

The Fermi surface tomography mathematical model is similar to computed tomography (CT), however given its development happened in parallel to mathematical advances in CT, mathematical descriptions that link both are missing from the literature. This section introduces the Fermi surface tomography forward model, and the one of the most successful and widely applied reconstruction algorithms, namely the Modified Cormack Method (MCM), describing the mathematical foundations and linking it to an operator form of the reconstruction, which can  open up the model for further analysis from the inverse problems community (analysis not presented in this work). Thus the description of the MCM presented here differs from the standard one in the Fermiology literature. We recommend the works of \cite{cormack_1963, cormack_1964, modified_cormack_method} for the more standard description of the MCM.

Firstly, the forward model of the Fermi surface tomography is the shown in Equation \ref{eq:Ander_forward} as

\begin{equation}
    \mathcal{R}\bar{\rho}^{2\gamma} = \bar{f} + \tilde{e}\label{eq:Ander_forward}
\end{equation}
where $\bar{f}$ is the measurement as a vector containing the positive half of our normalized projections as ${\bar{f}} \in \mathbb{R}^{L}$, where $L=L_p \cdot \Phi$ being the number of measured pixels $L_p$ in each measured projection $\Phi$ and $\bar{\rho}^{2\gamma} \in \mathbb{R}^{2L_P\cdot 2L_p} $ is the vectorized reconstructed TPMD image, finally $\tilde{e}$ correspond to measurement noise. The operator $\mathcal{R}$ is the Radon transform. 

In practice, the measurements are 2D over $\Phi$ projections at various angles $\varphi$ of a sample's 3D TPMD. Each projection $f_\varphi$ will have shape $f_{\varphi} \in \mathbb{R}^{L_p \times L_p}$, but due to mirror symmetry, only one quadrant of the data will be needed, and the other 3 quadrants will be symmetrically folded and averaged to reduce measurement noise, and finally normalized against the first projection. Given the operator is independent per slice of the 2D measurement $f_{\varphi} $, we define $f$ as a single row of $f_{\varphi}$, for every $\varphi$ and its vectorized version, $\bar{f}$. This allows us to define the operators in a more simple way, and highlights that the 3D-TPMD reconstruction process is comprised of a stack of independent 2D problems. 

In tomographic applications where the Radon transform is the operator, the pseudo-inverse $\mathcal{R}^\dagger$ (or reconstruction operator) is often represented by the Filtered Backprojection (FBP) algorithm. However, this is not the case in Fermi surface tomography, as the extremely limited number of projections causes the FBP algorithm to struggle. The MCM, by instead exploiting the inherent known symmetries of the reconstruction, produces superior results and thus has been widely applied. 

\subsubsection{The Modified Cormack Method}
The MCM algorithm can be written it in operator notation, as:
\begin{equation}
    \bar{\rho}^{2\gamma}=\mathcal{R}^\dagger \bar{f}=\mathcal{O}\mathcal{M}\bar{f}=\mathcal{O}\,
\mathcal{Z}\,
\mathcal{W}\,
\mathcal{S}^{-1}\,
\mathcal{P}^{-1}\,
\mathcal{C}\,
\bar{f},
\end{equation}
where each of the matrices represent a step of the MCM algorithm.  Firstly $\mathcal{C} \in \mathbb{R}^{M \times L}$ is a linear interpolator operator, in particular converting each row of the detector from a linear coordinate system sampled from the indices $t=\cos(\delta)$, with $\delta$ being a vector that linearly samples from $[0, \frac{\pi}{2}]$ of size $\Delta$, making $M=\Delta\cdot\Phi$. This step is often described in the MCM method as a projection of the measurement into the domain of the unit circle with polar coordinates. 

This radial sampling of the measurement allows to expand the data into polar Fourier series, defined by operator $\mathcal{P}\in \mathbb{R}^{M\times M}$. Let $N(t,\varphi) := (\mathcal{C}f)(t,\varphi)$ denote the projection data expressed in the polar coordinates produced by the interpolation step $\mathcal{C}$. This operator relies on an important observation, and prior information, of the crystal being measured and its symmetries. $G$ denotes the reciprocal lattice vector of the crystal being studied, and let $|G|$ denote its order, i.e.\ the fold of rotational symmetry. For a face-centred cubic (FCC) structure, this corresponds to $|G| = 4$ — which all of the samples studied in this work have. Because $N(t,\varphi)$ must be invariant under a rotation of $2\pi/|G|$, only harmonics of order $n \equiv 0
\mod{|G|}$ can appear in its Fourier expansion, so that

\begin{equation}
    N(t, \varphi) = \sum_{\substack{n = 0 \\ n \equiv 0 \,\mathrm{mod}\, |G|}}^{\infty} N_n(t) \cos(n\varphi).
    \label{Equation: Projection Data Polar Fourier Series}
\end{equation}

In operator form, $\mathcal{P}$ is thus a \emph{symmetry-truncated} Fourier transform: it maps $\mathcal{C}\bar{f}$ onto the coefficients $N_n(t)$ at orders $n \equiv 0\mod{|G|}$ only, discarding all other harmonics a priori. Having obtained the radial harmonic profiles $N_n(t)$, each is further expanded as a sine series built from Chebyshev polynomials as

\begin{equation}
    N_n(t) = 2 \sum_{m=0}^{\Delta} \tilde{a}_n^m \sin\left[(2m + 1) \cos^{-1}(t) \right] 
    \label{Equation: Chebyshev coefficients2}.
\end{equation}

\noindent Equation \ref{Equation: Chebyshev coefficients2} can be written in matrix form as $\mathcal{S} = I \otimes \mathcal{S}^\varphi$,  
where is $I\in \mathbb{R}^{\Phi \times \Phi}$ the identity matrix and $\mathcal{S}^\varphi \in \mathbb{R}^{\Delta \times \Delta}$ is\footnote{For equation simplicity this works uses the convention of matrix elements starting at index zero, i.e. the first element of a matrix is in index $i=0, j=0$.}

\begin{equation}
    \mathcal{S}^{\varphi}_{ij} = 2\sin\!\left[ (2j+1)\cos^{-1}\frac{i}{\Delta} \right].
\end{equation}

\noindent We can thus define the process of obtaining Chebyshev coefficients from measured projections as 
\begin{equation}
    \bar{a}=\mathcal{A}\bar{f}=\mathcal{S}^{-1}\mathcal{P}^{-1}\mathcal{C}\bar{f}.
\end{equation}

\noindent This operator $\mathcal{A}$ is useful later in this work to generate synthetic data.

Before the coefficients $\mathbf{a}_n$ are used to synthesize the
density, they are weighted elementwise with diagonal matrix $\mathcal{W}\in \mathbb{R}^{M\times M}$ as:

\begin{equation}
\mathcal{W}_{ij} =
\begin{cases}
2 (j\mod \Delta) + \lfloor \frac{j}{\Delta} \rfloor |G| + 1, & i = j \\
0, & i \neq j
\end{cases}
\end{equation}

The following step of the MCM reconstruction is to map the Chebyshev coefficients into the radial density functions, $\rho \in \mathbb{R}^{L\times1}$, using Zernike polynomials, which can be represented as 

\begin{equation}
R_n^m(\omega) = \sum_{k=0}^{\frac{n-m}{2}} \frac{(-1)^k \, (n-k)!}{k! \left( \frac{n+m}{2} - k \right)! \left( \frac{n-m}{2} - k \right)!} \, \omega^{n-2k},
\end{equation}
being $\omega$ the radial distance over the unit circle. The construction of the Zernike matrix $\mathcal{Z}\in \mathbb{R}^{L\times M}$ can be summarized as $\mathcal{Z}=I \otimes\mathcal{Z}^n$, with an identity matrix $I\in \mathbb{R}^{\Phi \times \Phi}$ and a series of $\mathcal{Z}^n \in \mathbb{R}^{L_p \times \Delta}$, for $n \in \{r|G| : r \in \{0, 1, \dots, \Phi\}\}$. Each sub-matrix is then built by computing the Zernike coefficients at the radial density values $w_i\in[0, 1, \dots,L]$, with a different matrix for each value of $n$, as

\begin{equation}
    \mathcal{Z}^n_{i,j}=R_n^j(w_i/L).
\end{equation}

This recovers the operator notation of the Modified Cormack Method:
\begin{equation}
\boxed{
    {\rho} = \mathcal{M}\bar{f}=\mathcal{Z}\,\mathcal{W}\mathcal{S}^{-1}
\mathcal{P}^{-1}\mathcal{C}\bar{f}.
}
\label{Methods: rho radial density function}
\end{equation}

The resulting radial density functions, $\rho$, can be converted to the density $\bar\rho^{2\gamma}(p,\theta)$
in the polar plane. Each channel $\rho_n(p)$ is modulated by
$\cos(n\theta)$ and summed:
\begin{equation}
    \rho^{2\gamma}_P=\rho^{2\gamma}(p,\theta) = \sum_{\substack{n = 0 \\ n \equiv 0 \,\mathrm{mod}\, |G|}}^{\infty}
    \rho_n(p)\cos(n\theta).\label{Methods:MCM_TPMD_Equation}
\end{equation}
The Equation \ref{Methods:MCM_TPMD_Equation} reconstruction step is illustrated in Appendix \ref{Appendix:MCM_Matrix}. This recombination step defines the operator $\mathcal{O}$, mapping ${\rho}$ onto $\bar\rho^{2\gamma}$, and gives the full
modified Cormack method as originally stated:
\begin{equation}
\boxed{
    \bar{\rho}^{2\gamma}_P = \mathcal{O}\,\mathcal{Z}\,\mathcal{W}
    \mathcal{S}^{-1}\,\mathcal{P}^{-1}\,\mathcal{C}\,f
}
    \label{Methods: rho_2gamma in Polar Space}
\end{equation}

While in Fermiology this is often the final step, it is of interest to also define the TPMD in Euclidean form ($\rho_E ^{2\gamma}$), as shown in Equation \ref{Equation:TPMD_Euclidean_Equation}, since the standard CNN architectures used in this work are more naturally suited to a Euclidean grid.

\begin{equation}
    \rho^{2\gamma}_{E} = \rho^{2\gamma}(p_x,p_z) = \sum_{\substack{n = 0 \\ n \equiv 0 \,\mathrm{mod}\, |G|}}^{\infty} \rho_n\left(\sqrt{p_x^2+p_z^2}\right) \cos\left(n \cdot \arctan\left(\frac{p_x}{p_z}\right)\right).
    \label{Equation:TPMD_Euclidean_Equation}
\end{equation}

In operator form, we express this transformation into Euclidean space using operator $\mathcal{E}$ as:

\begin{equation}
\boxed{
    \bar{\rho}^{2\gamma}_E = \mathcal{E}\,\mathcal{Z}\,\mathcal{W}
    \mathcal{S}^{-1}\,\mathcal{P}^{-1}\,\mathcal{C}\,f
}
\label{Methods: rho_2gamma in Euclidean Space}
\end{equation}
\FloatBarrier

\subsection{DeepCormack \label{Methods: DeepCormack Design}}
\begin{figure}[htp!]
  \centering
  \includegraphics[width=0.98\textwidth]{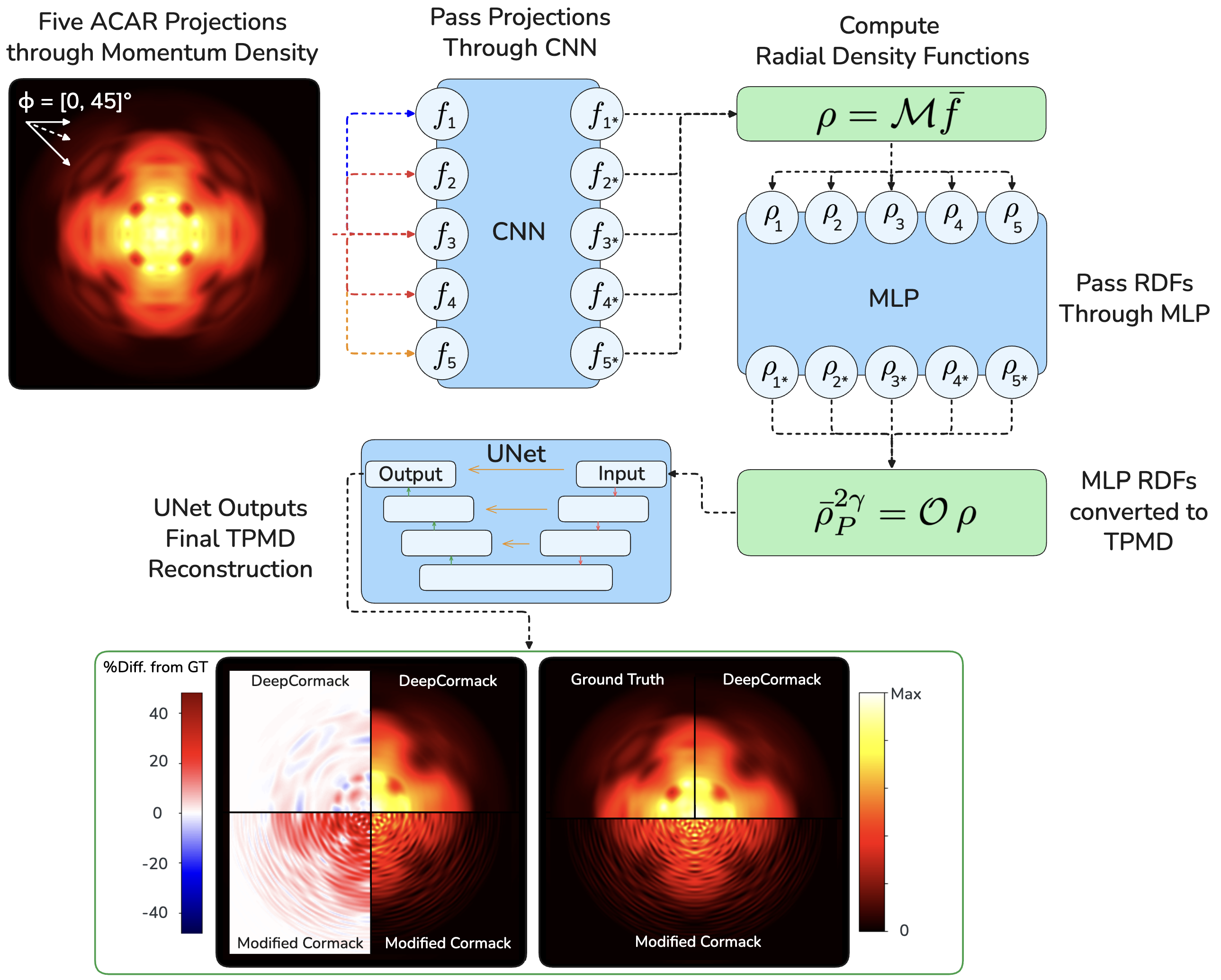}
  \caption{\footnotesize{\textit{Workflow of the {\fontfamily{lmtt}\selectfont DeepCormack} Method ($10$M Counts), illustrating which stages of the Modified Cormack Method (green) neural networks are added (blue).}}}
  \label{fig:DeepCormack_Workflow}
\end{figure}

\par
This section describes our proposed data-driven method, enhancing the MCM algorithm with deep neural networks. In general, {\fontfamily{lmtt}\selectfont DeepCormack} integrates up to three neural networks at different key stages of the modified Cormack method, as shown in Figure \ref{fig:DeepCormack_Workflow}. The process works as follows: 2D-ACAR measures projections of a sample's momentum density at different angles.

The MCM reconstructs the TPMD at fixed detector row slice-by-slice, and {\fontfamily{lmtt}\selectfont DeepCormack} proceeds similarly. However, the normalized projections are first passed through a convolutional neural network (CNN) (details in section \ref{Method: 1DCNN}) in order to denoise and deconvolute them. Equation \ref{Methods: rho radial density function} is then applied to recover the corresponding $\sim 5$ radial density functions, which are subsequently passed through a fully connected multilayer perceptron (MLP) (details in section \ref{Method: MLP}). The MLP-outputted radial density functions are passed through Equation \ref{Methods:MCM_TPMD_Equation} to reconstruct the TPMD $\rho ^ {2 \gamma}$. This reconstruction is finally passed through a UNet (details in section \ref{Method: UNet}).

In this work, we explore several possible configurations of this  {\fontfamily{lmtt}\selectfont DeepCormack} structure explained in more detail in the Results section.

\subsubsection{1D CNN \label{Method: 1DCNN}}
\begin{figure}[htp!]
  \centering
  \includegraphics[width=0.98\textwidth]{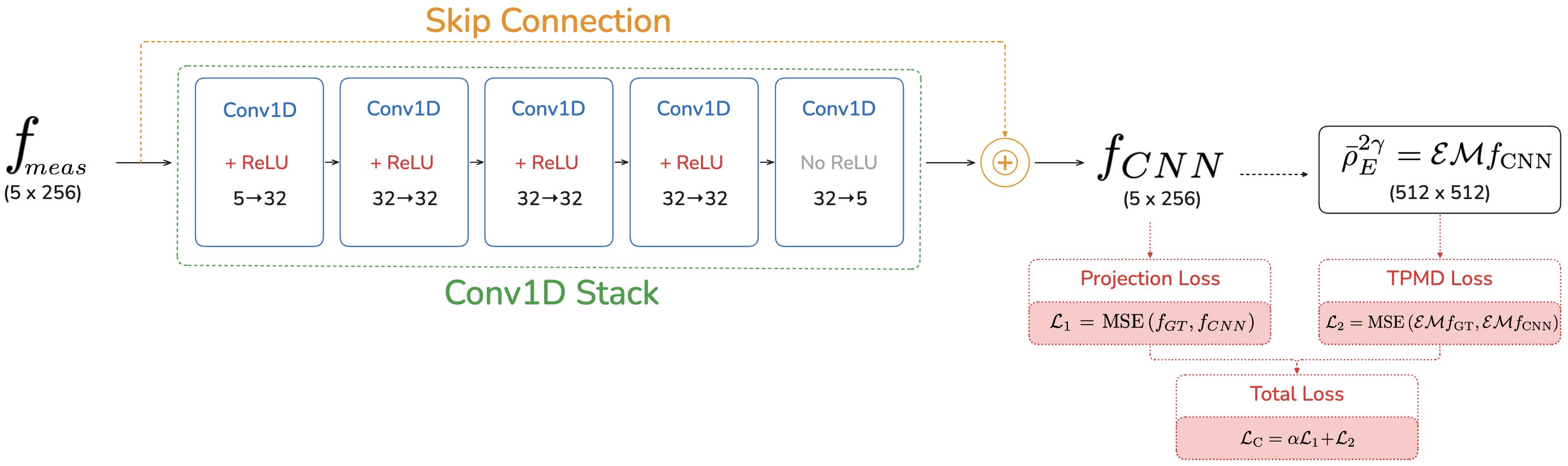}
  \caption{\footnotesize{\textit{Workflow of the The 1D CNN.}}}
  \label{fig:1D_CNN_Workflow}
\end{figure}

The first stage, shown in Figure \ref{fig:1D_CNN_Workflow}, employs a one-dimensional residual convolutional neural network (CNN) to map noisy measured projections $(f_{meas}=f+\tilde{e})$ to their ideal counterparts $({f}_{GT}=f)$. The network accepts as input $\Phi$ ($5$ in our work) angular projections, each of length $L_p$ ($256$ in ours), arranged as a multi-channel 1D signal of shape $(\Phi, L_p)$. The architecture consists of a stack of five one dimensional convolutional layers with kernel size $3$ and unit padding, with the first four layers projecting into $32$ hidden channels and the final layer projecting back to $5$ output channels, preserving the input dimensionality. ReLU activations are applied between all intermediate layers. A global residual connection is added from the input directly to the output, such that the network learns a correction $\Delta$ rather than the full mapping, i.e. ${{{f}}}_{CNN} = {{f}}_\text{meas} + \text{CNN}({{f}}_\text{meas})$. The network is trained by minimising the loss comprising a direct mean-squared error (MSE) between predicted and ideal projections ($\mathcal{L}_{1} = \text{MSE} \left({{f}}_{GT}, {{f}}_{CNN} \right) $), and an auxiliary MSE between the ground truth TPMD and CNN-denoised TPMD reconstruction in Euclidian coordinates ($\mathcal{L}_{2} = \text{MSE} \left(\mathcal{EM}{f}_\text{GT}, \mathcal{EM}{f}_\text{CNN} \right)$). The full loss term is constructed as $\mathcal{L}_\text{C} = \alpha \mathcal{L}_1 + \mathcal{L}_2$. We found $\alpha =10$ empirically the best value. This allows the model to denoise the projections, while ensuring the final results are still valid, given that small changes in the measurements can cause large changes in the final image. The network is optimised using the Adam optimiser with a learning rate of $10^{-3}$.

\subsubsection{MLP \label{Method: MLP}}
\begin{figure}[htp!]
  \centering
  \includegraphics[width=0.98\textwidth]{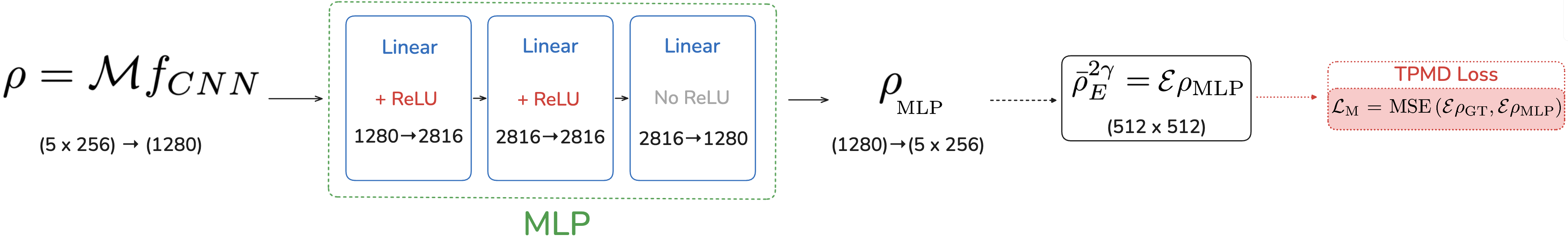}
  \caption{\footnotesize{\textit{Workflow of the The MLP.}}}
  \label{fig:MLP_Workflow}
\end{figure}
The second stage employs a fully connected multilayer perceptron (MLP) to refine the radial density function representation of the CNN output, as shown in Figure \ref{fig:MLP_Workflow}. The input to the MLP is a tensor of shape $(\Phi,L_p)$ representing the $L_p = 256$ radial grid points each described by $\Phi=5$ Cormack expansion coefficients, which is flattened to a vector of dimension $\Phi\cdot L_p=1280$. The network consists of two hidden layers of width $11 \times L_p = 2816$ with ReLU activations, followed by a linear output layer projecting back to the input dimensionality. The output is reshaped to $(\Phi,L_p)$ to match the input. The CNN is frozen, and the MLP receives as input the CNN-denoised projections passed through the MCM to get ${\rho}$. The network is supervised entirely in reconstruction space: the loss is the MSE between the Euclidean 2D TPMD from the MLP output and the ground truth ($\mathcal{L}_\text{M} = \text{MSE} \left(\mathcal{E}{\rho}_\text{GT}, \mathcal{E}{\rho}_\text{MLP} \right)$). This formulation also encourages the MLP to recover features in the TPMD that are lost due to the limited number of projections available in experimental ACAR. The MLP is optimised with the Adam optimiser at a learning rate of $10^{-3}$.

\subsubsection{UNet \label{Method: UNet}}
\begin{figure}[htp!]
  \centering
  \includegraphics[width=0.98\textwidth]{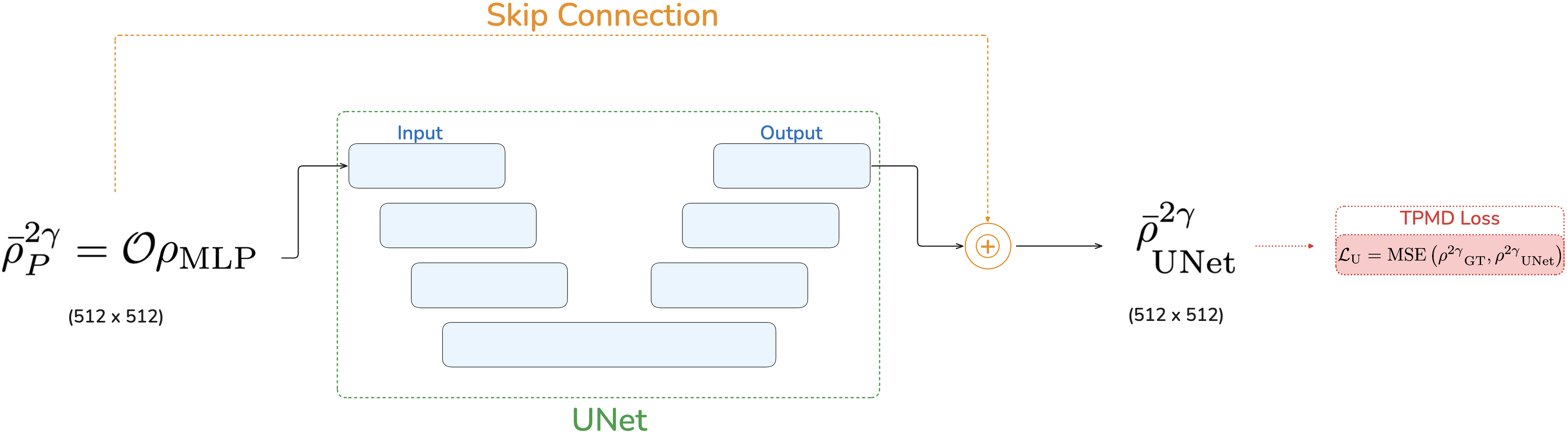}
  \caption{\footnotesize{\textit{Workflow of the The UNet.}}}
  \label{fig:UNet_Workflow}
\end{figure}
The third stage is a UNet architecture (Figure \ref{fig:UNet_Workflow}) based on \cite{UNet_Structure} to learn residual corrections to the two-dimensional TPMD reconstruction. The projections are passed through the CNN, converted to radial density functions ${\rho}$, and inputted to the MLP. The output is reconstructed into a quadrant of the 2D image ${\rho}^{2\gamma}_P$ of shape $(1, L_p, L_p)$ via the final step of the MCM, and becomes the input of the UNet.

The encoder comprises three convolutional blocks with channel progressions $[1, 64, 64, 64]$, $[64,$ $128,$ $128]$, and $[128, 256, 256]$ respectively, each followed by $2\times2$ max-pooling for spatial downsampling. The bottleneck processes features at $[256, 512, 512]$ channels with a spatial dropout rate of $0.1$. The decoder mirrors the encoder using transposed convolutions for upsampling, with skip connections concatenating encoder feature maps to the corresponding decoder inputs at each resolution level, and channel progressions $[512, 256, 256]$, $[256, 128, 128]$, and $[128, 64, 64]$. Each convolutional block uses Group Normalisation and ReLU activations. A final $1\times1$ convolution projects to a single output channel. A global residual connection adds the input image to the network output, enforcing that the UNet learns a residual correction such that ${\rho}^{2 \gamma} = \text{ReLU}(\rho^{2 \gamma} _\text{pred} + \text{UNet}(\rho^{2 \gamma} _\text{pred}))$, where the ReLU enforces physical non-negativity of the momentum distribution. The network is trained by minimising the MSE between the corrected output and the ground-truth TPMD image ($\mathcal{L}_\text{U} = \text{MSE} \left({\rho^{2\gamma}}_\text{GT}, {\rho^{2\gamma}}_\text{UNet} \right)$), using the Adam optimiser at a learning rate of $10^{-4}$.

\subsubsection{FiLM Conditioning \label{Method: FiLM Conditioning}}
A UNet was also tested with and without Feature-wise Linear Modulation (FiLM) \cite{FiLM_Original_Paper, FiLM_Neurips}. FiLM learns functions $\gamma(z)$ and $\beta(z)$ to modulate a neural network's intermediate feature map $\mathbf{x}$ via a learned affine transformation:

$$
\text{FiLM}(\mathbf{x} \mid \gamma, \beta) = (1 + \gamma(z)) \cdot \mathbf{x} + \beta(z),
$$
\noindent making it possible to condition the network on a signal $z$ without modifying the weights themselves. In our case, $z$ is taken as the log-normalised count level averaged across the simulated measurement projections, providing the UNet with an explicit measure of the Poisson noise in the data. 

For the UNet, FiLM is applied to the output of each convolutional block (i.e., after normalisation and activation), rather than between normalisation and the nonlinearity as in the original formulation. $z$ is first encoded via a sinusoidal embedding and refined by a shared two-layer SiLU MLP, whose output is distributed to a single linear projection per block, producing per-channel $(\gamma, \beta)$ pairs at each encoder, latent, and decoder stage. FiLM parameters are zero-initialised, such that at the start of training $\gamma = 0$, $\beta = 0$, making the transformation an identity map.

\subsection{Data Simulation \label{Methods: Data Simulation}}

As experimental ACAR measurements can take months, constructing 3D TPMD training datasets of the size typically required for supervised deep learning via experimental means alone would be impractical, as it takes 3-4 months to measure a single materials. Computational approaches such as density functional theory (DFT) could in principle provide ground-truth momentum densities from which projections could be taken through them and realistic experimental conditions simulated. However, DFT calculations are themselves rather intensive to compute both in terms of human time (parameter setting, convergence testing etc.) as well as resources (computing times can be of the order of hours, and the densities themselves can require storage at the GB scale). As no publicly available momentum density database currently exists, a key step for making supervised learning feasible requires a data generation method for sufficiently diverse and realistic synthetic 3D TPMDs that continuously evolve slice-by-slice. This would allow the detector resolution to be simulated as a 2D Gaussian convolution, noise via Poisson statistics, and the momentum sampling function (MSF) — a necessary correction due to the ACAR detector geometry, as only photons emitted with perfect collinearity have 100\% probability to both hit the detectors — can be applied as a normalized tent function.

Our proposed data generation approach is applied on a reference momentum density for copper, and consists of two independent steps: first, plausible central slices are generated, using a combination of singular value decomposition (SVD) on the copper Chebyshev coefficients and varying the singular values. Dynamic mode decomposition (DMD) \cite{DMD_BruntonKutz} is then applied, which learns the underlying dynamics of how the copper momentum density evolves continuously from one slice to the next.

Any number of central slice TPMDs can be generated using SVD, then continuously evolved into a full 3D TPMD using DMD. This will act as our ground truth for the dataset. As outlined in Section \ref{Method: Making Realisting Projections}, a corresponding experimental dataset can be simulated by taking $5$ projections through the ground truth volumes and applying realistic degradation to the data (detector resolution and geometry effects, noise etc). 

Taking $20$ ideal projections of the DFT copper momentum density between $\varphi \in [0, 45]$ degrees preserves sufficient information when reconstructed via the MCM (Equation \ref{Methods: rho_2gamma in Polar Space}) to act as an approximation for the ground truth. This allows our data generation method to operate entirely on the extracted Chebyshev coefficients, as defined in Equation \ref{Equation: Chebyshev coefficients2}, for some projection slice $l = p_y \in [0, L_p-1],$ $L_p=256,$ that is reconstructed from copper: $\tilde{a}_{n \hspace{0.05cm}, \hspace{0.05cm} l}^m$. Here, $m \in [0, \Delta-1],$ $ \Delta = 180$ are the number of angular degrees of freedom when parametrising the projections along the unit circle, and $n \in [0, \Phi-1]$ denotes the projection index, where $\Phi=20,$ are the number of ideal projections used to approximate the ground truth copper momentum density with the MCM. For some projection $\Phi,$  we define the matrix $\left[ \mathbf{\tilde{a}_\Phi} \right]_{lm} = \tilde{a}_{\Phi \hspace{0.05cm}, \hspace{0.05cm} l}^m$, where $\mathbf{\tilde{a}}_\Phi \in \mathbb{R}^{\Delta \times L_p}$. When generating the synthetic data, we model each of the Chebyshev coefficients independently via their projection channel by fixing $n$, as they are tied to separate $\rho$'s when reconstructed. The label will be omitted hereafter for notational clarity.

\subsubsection{Generating Central Slices \label{Method: SVD}}
The MCM, at the step shown in Equation \ref{Equation: Chebyshev coefficients2}, decomposes each projection into a set of Chebyshev coefficients to carry out the reconstruction. These are the source values used for our generative process of central slices.

For a fixed projection, truncated SVD is applied across the angular channel over all unique slice indices, reducing the dimensionality to a level that maintains approximately $\sim 90 \%$ of the variance. This yields an independent lower dimensional latent space per projection channel, each of which is modelled by fitting a multivariate Gaussian to allow sampling new points. When lifted back to coefficient space, the result is a new set of Chebyshev coefficients that can proceed through the rest of the MCM (Equation \ref{Methods: rho_2gamma in Polar Space}), and after some minor tuning will give a realistic central TPMD slice.

More precisely, for $\mathbf{\tilde{a}}$ previously defined in Section \ref{Methods: Data Simulation}, let $\mathbf{\tilde{a}}_{l} \in \mathbb{R}^{\Delta},$ denote the copper Chebyshev coefficient state at a fixed projection $n$ at a given slice $l$. To construct the latent space and sample a new set $\hat{\mathbf{a}} \in \mathbb{R}^{\Delta},$ we begin by constructing the snapshot matrix using the copper values:

\begin{equation}
\mathbf{X} =
\begin{bmatrix}
\mathbf{\tilde{a}}_0 & \mathbf{\tilde{a}}_1 & \cdots & \mathbf{\tilde{a}}_{L_p -1}
\end{bmatrix}
\in \mathbb{R}^{\Delta \times L_P}.
\label{Equation: Chebyshev_Matrix_fixedN}
\end{equation}

\noindent The temporal\footnote{Note that time is not a variable in our work, instead the Chebyshev coefficient variation over slices are the variable. However, for consistency with the DMD and Koopman operator literature, we call this time.} mean is defined as
$$
\bar{\mathbf{a}} = \frac{1}{L_p} \sum_{l=0}^{L_p-1} \mathbf{\tilde{a}}_l \in \mathbb{R}^{\Delta},
$$

\noindent allowing us to get the mean-centered data as:

$$
\mathbf{X}_c = \mathbf{X} - \bar{\mathbf{a}} \mathbf{1}^\top,
$$

\noindent where \(\mathbf{1} \in \mathbb{R}^{L_p}\) is a vector of ones.

We compute the singular value decomposition (SVD) of the centered snapshot matrix:
$$
\mathbf{X}_c = \mathbf{U} \boldsymbol{\Sigma} \mathbf{V}^\top,
$$

\noindent where $\mathbf{U} \in \mathbb{R}^{\Delta \times \Delta}$ contains the orthonormal spatial modes, $\boldsymbol{\Sigma} \in \mathbb{R}^{\Delta \times L_p}$ is a diagonal matrix of singular values, and $\mathbf{V} \in \mathbb{R}^{L_p \times L_p}$ contains temporal coefficients. We retain the first $K \ll \Delta$ dominant modes, yielding the truncated SVD:

\begin{equation}
  \widetilde{\mathbf{X}}_c \approx \mathbf{U}_k \boldsymbol{\Sigma}_k \mathbf{V}_k^\top.
  \label{Equation: Truncated SVD}
\end{equation}

\noindent $\mathbf{U}_k \in \mathbb{R}^{\Delta \times K}$ defines a reduced spatial basis. In this step, $K=15$ $(\Delta/K = 12)$ was found to construct reasonable TPMD features, capturing $\sim 88 \%$ of the total variance across slices. A Gaussian is fitted to the latent coordinates $\boldsymbol{\Sigma}_k \mathbf{V}_k^\top \in \mathbb{R}^{K \times L_p}$ of the $L_p = 256$ copper slices, a new latent vector $\hat{\mathbf{z}}$ is drawn, and projected back to coefficient space to give a synthetic Chebyshev coefficient $\mathbf{\hat{a}} \in \mathbb{R}^{\Delta}$ by:

$$
  \hat{\mathbf{z}} \sim \mathcal{N}(\boldsymbol{\mu_k},\,\boldsymbol{\Lambda_k}),
    \hspace{1cm}
  \hat{\mathbf{a}}
  = \mathbf{U}_k\,\hat{\mathbf{z}} + \bar{\mathbf{a}}
  \in \mathbb{R}^{\Delta},
$$

\noindent where $\mu_k$ and $\Lambda_k$ are the sample mean and full covariance matrix of the latent coordinates respectively. Repeating this process for all $\Phi$ channels (with the projection channel $n$ reinstated) yields the new central-slice coefficient matrix
$$
  \hat{\mathbf{A}}_{\mathrm{l=0}}
  =
  \bigl[
    \hat{\mathbf{a}}_{\mathrm{0}},\;\dots,\;
    \hat{\mathbf{a}}_{\mathrm{\Phi-1}}
  \bigr]
  \in \mathbb{R}^{\Delta \times \Phi}.
$$

Equation \ref{Methods: rho radial density function} can subsequently be applied to recover the corresponding $\rho$ and Equation \ref{Methods: rho_2gamma in Polar Space} for the reconstructed $\rho^{2 \gamma}_P$.

It was found that applying corrections to the $\rho$ generated from $\hat{\mathbf{A}}_{\mathrm{l=0}}$ improved the overall quality and consistency of the SVD sampling method. First, the isotropic term $\rho_0$ (being the most sensitive and frequently producing poor-quality TPMDs when sampled via SVD) is replaced with the isotropic term drawn from the original copper dataset at a slice index in the range $l \in [0,133]$. 1D Gaussian smoothing is applied along the radial axis of the higher-order density terms $\rho_{n>0}$ to suppress unphysical and overly exotic features that arise from SVD sampling but are not characteristic of real TPMDs. All $\rho_n$ functions are normalised such that their peak intensities remain within a $\sim 10 \%$ range observed in the corresponding copper reference functions. Finally, a physicality constraint is then enforced: the total radial density $\sum_n \rho_n$ must remain non-negative everywhere. Passing these corrected $\rho_n$ functions through the final MCM reconstruction step yields the synthetic ground truth TPMD slice to be used in the next stage of the data generation method.

Since corrections have been applied to the $\rho_n$ functions, evolving them into full 3D TPMDs will require recomputing the Chebyshev coefficients for each synthetic central slice. This is done by applying the Radon transform at 20 equally spaced projection angles $\varphi \in [0^{\circ}, 45^{\circ}]$ and recovering the corresponding (corrected) Chebyshev coefficients via the MCM (Equation \ref{Methods: rho radial density function}).

\subsubsection{Evolving the central slice into 3D TPMDs \label{Method: DMD}}

Once feasible generated central slices are created, 3D versions of them need to be derived. In order to extrapolate the central slice in a manner coherent with TPMD structure, we propose a dynamic mode decomposition (DMD) model \cite{DMD_BruntonKutz}. This is trained per-projection channel on the original copper Chebyshev coefficients $\mathbf{\tilde{a}}$. DMD is a particularly powerful method for this: it allows the slice-to-slice non-linear dynamics of the Chebyshev coefficients to be simulated by assuming they evolve linearly in a latent space. It achieves this by computing an approximate Koopman operator for each $n$, such that the synthetically generated set of central-slice Chebyshev coefficients $\hat{\mathbf{A}}_{\mathrm{l=0}}$ can be evolved into a full set $\hat{\mathbf{A}} \in \mathbb{R}^{L_p \times \Delta \times \Phi}$. Passing each of the evolved Chebyshev coefficients through the MCM as $\mathcal{A}\bar{f}$ in Equation \ref{Methods: rho radial density function}, the result is a continuously evolving 3D TPMD.

In order to compute the DMD model, we Continue from the truncated SVD for copper in Equation \ref{Equation: Truncated SVD}, for a fixed projection channel $n$, we now retain the first $K=24$ ($M/K = 7.5$) dominant modes as it was found to give a good trade-off between diversity and stable volume evolution. Each Chebyshev slice snapshot $(\tilde{\mathbf{a}}_l)$ from Equation \ref{Equation: Chebyshev_Matrix_fixedN} is projected onto the reduced basis as
$$
\mathbf{z}_l = \mathbf{U}_k^\top (\tilde{\mathbf{a}}_l - \bar{\mathbf{a}}) \in \mathbb{R}^{K}.
$$

\noindent Allowing us to define the time-shifted latent snapshot matrices:
\[
\mathbf{Z}
=
\begin{bmatrix}
\mathbf{z}_0 & \mathbf{z}_1 & \cdots & \mathbf{z}_{L_p-2}
\end{bmatrix}
\in \mathbb{R}^{K \times (L_p-1)},
\]
\[
\mathbf{Z}'
=
\begin{bmatrix}
\mathbf{z}_1 & \mathbf{z}_2 & \cdots & \mathbf{z}_{L_p-1}
\end{bmatrix}
\in \mathbb{R}^{K \times (L_p-1)}.
\]

\noindent We assume linear evolution in latent space:
\[
\mathbf{z}_{l+1} = \mathbf{\mathcal{K}} \mathbf{z}_l,
\]
\noindent where \(\mathcal{K} \in \mathbb{R}^{K \times K}\) is a finite-dimensional approximation of the Koopman operator, and is obtained by solving the least-squares problem
\[
\min_{\mathcal{K}} \| \mathbf{Z}' - \mathcal{K} \mathbf{Z} \|_F,
\]
\noindent where $||\cdot||_F$ is the Frobenius norm. The closed-form solution is
\[
\mathcal{K} = \mathbf{Z}' \mathbf{Z}^{\dagger},
\]
\noindent with \(\mathbf{Z}^{\dagger}\) denoting the Moore--Penrose pseudoinverse.

Once the Koopman operator $\mathcal{K}$ is pre-computed in latent space, the generated central slice for the projection channel $\mathbf{\hat{a}}_l$ can be rolled out to produce slice $\hat{a}_{l+1}$ via:

\begin{equation}
{\hat{\mathbf{a}}}_{l+1}
=
\mathbf{U}_k \mathcal{K} \mathbf{U}_k^\top (\hat{\mathbf{a}}_l - \bar{\mathbf{a}})
+ \bar{\mathbf{a}},
\label{Equation:DMD_evolution}
\end{equation}

\noindent resulting in the new Chebyshev snapshot matrix for projection channel $n$:

$$
\hat{\mathbf{A}}_n =
\begin{bmatrix}
\hat{\mathbf{a}}_0 & \hat{\mathbf{a}}_1 & \cdots & \hat{\mathbf{a}}_{L_p-1}
\end{bmatrix}
\in \mathbb{R}^{\Delta \times L_p}.
$$

Repeating this process for all $\Phi$ projection channels — each with their own unique Koopman operator \(\mathcal{K}_n\) — will result in a full set of synthetic Chebyshev coefficients:

$$
\hat{\mathbf{A}} =
\begin{bmatrix}
\hat{\mathbf{A}}_0 ^T & \hat{\mathbf{A}}_1 ^T & \cdots & \hat{\mathbf{A}}_{\Phi-1} ^T
\end{bmatrix}
\in \mathbb{R}^{ L_p \times\Delta \times \Phi}.
$$

Separately passing each $\hat{\mathbf{A}}_l = [\mathcal{A}f]_{\text{synthetic}}$ from $l = 0$ to $L_p-1$ into Equation \ref{Methods: rho_2gamma in Polar Space} will result in a full 3D TPMD which continuously evolves. Each time a new central slice is constructed with SVD and evolved with DMD, a unique 3D TPMD is created allowing countless samples to be generated for training.

The effect of taking the copper central slice Chebyshev coefficents and applying DMD to continuously evolve them with the copper-derived Koopman operators is shown in Appendix \ref{Appendix:DMD_3D_TPMD}, where the Cu ground truth and DMD Fermi surfaces are compared.

\subsubsection{Simulating Experimental Measurement Conditions \label{Method: Making Realisting Projections}}

\begin{figure}[htp!]
  \centering
  \includegraphics[width=0.98\textwidth]{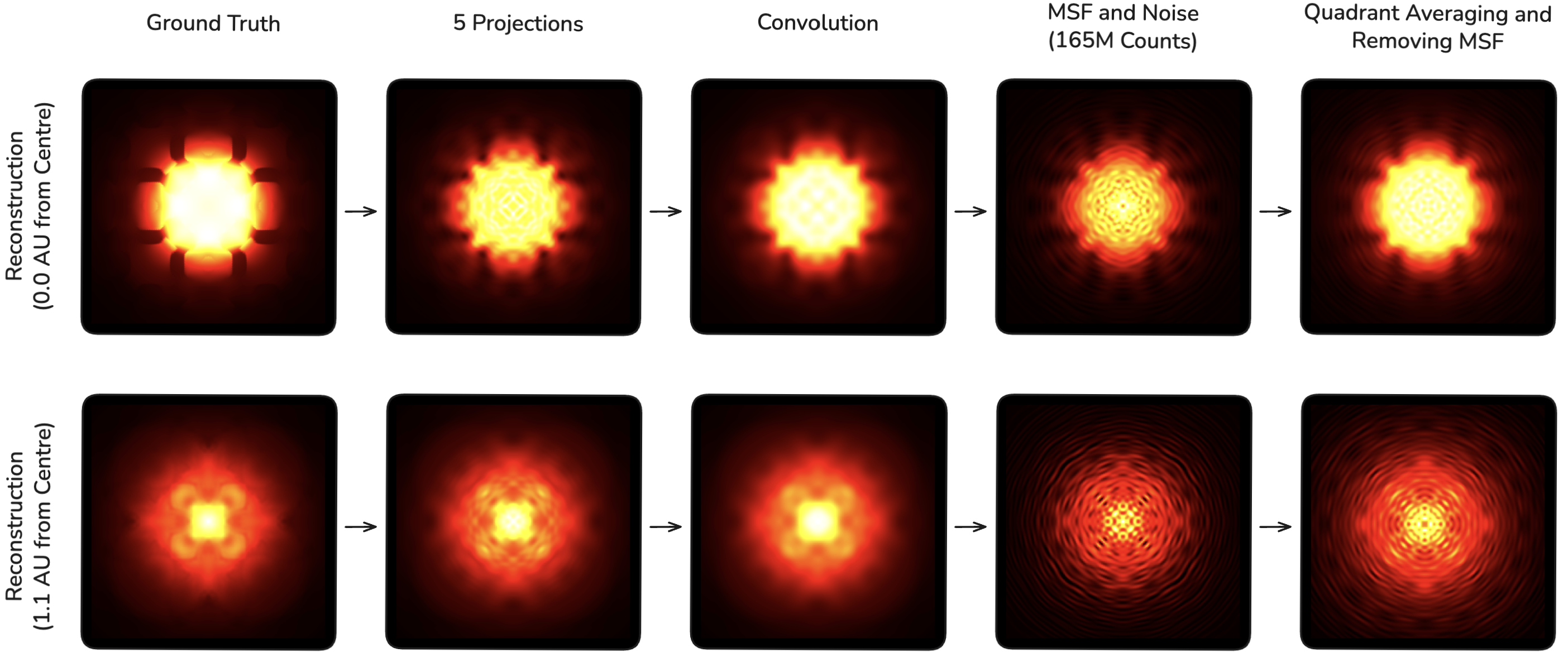}
  \caption{\footnotesize{\textit{Step-wise simulation of experimental conditions on projections through the ZrZn$_2$ momentum density. The resulting image degradation from the modified Cormack method is shown for a slice through the centre and 1.1 au outwards. From (i) the ground truth, the procedure includes (ii) taking $5$ projections of size $6.57 \times 6.57$ au between $\varphi \in [0,\pi/4]$ degrees through the ground truth, (iii) applying an elliptical Gaussian to model detector resolution with $(\sigma_x, \hspace{0.1cm} \sigma_y) = (0.110, \hspace{0.1cm} 0.137)$ au, (iv) multiplying by a normalized momentum sampling function (MSF) before adding Poisson noise to each projection at $165$M counts, (v) symmetrically averaging the quadrants of the projections before dividing by the same MSF.}}}
  \label{fig:Simulating_Meas_Projections}
\end{figure}

A final step of generating realistic data for supervised ML is to be able to produce noise similar to the one present in experimental conditions at different count levels. Figure \ref{fig:Simulating_Meas_Projections} shows our simulated noise pipeline and its effect in MCM reconstructed slices, for both a central slice and a slice $1.1$ au away from it.

For a synthetic 3D TPMD volume, (i) five 2D projections are computed at angles $\varphi \in [0^{\circ}, 45^{\circ}]$. (ii) Detector resolution effects ($0.110 \times 0.137$ a.u. smearing along the projection axis $p_x$ and $p_y$), (iii) Poisson counting noise and momentum sampling function (MSF) effects are applied to each projection in order to simulate realistic experimental conditions at the desired count rates. The MSF arises as a consequence of the detector geometry, where only electrons emitted with perfect collinearity have a $100\%$ probability of both registering a hit at either detector. This effect can be simulated by multiplying the projections with a normalized tent function. As the synthetic dataset is based on a momentum density for copper, it will retain the sample's four-fold symmetry, and (iv) each projection's quadrants can be averaged with one another before removing the MSF contribution. MCM reconstruction is then applied to these degraded projections. This yields matched pairs of ground truth and simulated measurement projections, radial density functions, and 3D TPMD volumes for use in supervised training.

\subsection{Training data for {\fontfamily{lmtt}\selectfont DeepCormack}}
To reduce storage and memory requirements, only a subset of $p_y$-slices are retained from each volume. These are drawn randomly from the range $p_y \in [0, 83]$ out of 256 total slices, a region which accounts for approximately 90\% of the total counts in the copper reference data. Appendix \ref{Appendix:SynthDataGen} Figure \ref{AppendixFig:SynthTPMDsComparison} shows three slices taken through different synthetic 3D momentum densities that were generated using our method and used in training.

\section{Results \label{Results}}

In this section we present experiments of trained {\fontfamily{lmtt}\selectfont DeepCormack} models at different scenarios. Before showing model performance, we first validate our noise simulation approach to the simulation of experimental conditions at varying counts by comparing the Fermi surface degradation of ZrZn$_2$ with that of real experimental data  (\ref{Results: Validating Sim Approach}). Different {\fontfamily{lmtt}\selectfont DeepCormack} model configurations are then tested and compared using three different datasets with simulated measurement conditions at $200$M counts: the synthetically generated test data (DMD), the copper TPMD which the synthetic training data used as reference, and ZrZn$_2$ data (\ref{Results: Model Comparison}). This allows for testing the models' reconstruction capabilities on data that is within the distribution of the training data (the DMD data), data which has features and slice-by-slice structure within distribution of the training data but not directly seen during training (the copper data), and on data out of distribution (OOD) entirely (the ZrZn$_2$ data). 

In addition to higher image quality, one goal of {\fontfamily{lmtt}\selectfont DeepCormack} is to enable faster scanning time. The next experiment presented in Section \ref{Results: Lowering Counts} showcases the model's performance at lowering the photon counts measured (i.e. reducing acquisition time). This is tested with the same three datasets from the previous experiments, i.e. the DMD, copper, and ZrZn$_2$ data.

\subsection{Validating the Simulation Approach \label{Results: Validating Sim Approach}}

\begin{figure}[htp!]
  \centering  
  \includegraphics[width=0.98\textwidth]{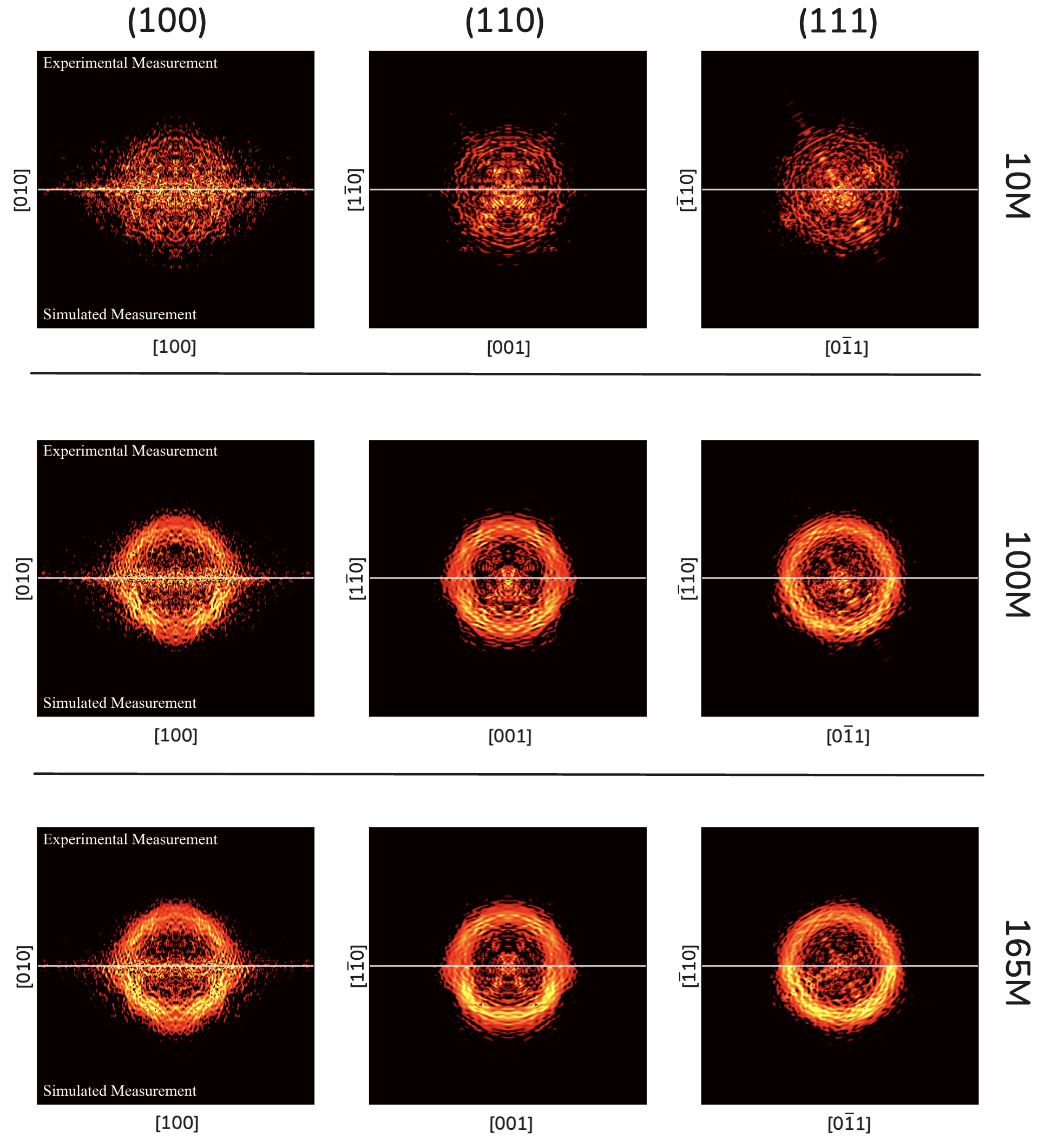}

  \caption{\footnotesize \textit{Gradient of the ZrZn$_2$ 3D TPMD reconstruction vs counts (165M, 100M, 10M), comparing the experimental measurement with the simulated measurement through the corresponding slice. The simulated measurement data used 60\% greater convolution and 40\% of the true counts (i.e. greater noise) to give comparable degradation quality to the actual experimental data.}}
  \label{AppendixFig:FSvsCounts}
\end{figure}

To validate the {\fontfamily{lmtt}\selectfont DeepCormack} framework on real data, the simulated experimental conditions must first be verified against real experimental measurements. This is to confirm that the synthetic data faithfully reproduces the degradation due to noise and smearing of real ACAR data across a range of count levels. 
To this end, we compare simulated and experimental ACAR measurements of ZrZn$_2$ across the range $10-165$M counts, using a DFT-calculated momentum density as the ground truth. Computing the gradient of the 3D TPMDs allows calibration quality to be assessed in TPMD space rather than by comparing the reconstructed images directly. The ACAR measurements had simulated measurement conditions applied as outlined in Section \ref{Method: Making Realisting Projections}, and repeated for data at $10$M, $100$M, and $165$M counts in order to provide direct comparison with real experimental data.

Using the nominal experimental parameters gave simulated data that was too smooth and insufficiently noisy, with degradation in Fermi space that was too mild relative to experiment. A better match was obtained by increasing the convolution widths by $+60\%$ to the reference values and using an effective count level of $40\%$ of the experimental total (e.g. $100$M experimental counts $\rightarrow$ $40$M simulated counts). With these adjusted parameters, the degradation matched the experimental reference well in Fermi surface space as shown in Figure \ref{AppendixFig:FSvsCounts}, and was similarly consistent in TPMD space (Appendix \ref{Appendix:SimExpVal} Figure \ref{AppendixFig:TPMDvsCounts}).  This mismatch is probably explained by the fact that there are other possible factors producing noise in real measurements that our first order approximation of the noise does not capture.

This validates the proposed forward model and approximation of the measurement noise, allowing us to generate synthetic measurement data with the correct noise characteristics, a necessary step for the supervised learning proposed in this work, that uses the synthetic TPMDs explained in Section \ref{Methods: Data Simulation}. The rest of the experiments use this simulated noise model for training.

\subsection{Model Comparison at Standard Counts \label{Results: Model Comparison}}

Several different {\fontfamily{lmtt}\selectfont DeepCormack} model configurations (defined below) were evaluated across three datasets to assess their reconstruction performance and ability to generalize. The DMD test set shares the same distribution as the training data; the copper (Cu) dataset represents a real TPMD, but is still within reasonable distribution of the training data, as it was used as the reference momentum density for the DMD generation; and the ZrZn$_2$ dataset introduces unique and unseen TPMD features, providing a useful test of OOD generalization. Due to the SNR of the 2D-ACAR technique, a standard experimental measurement will capture around $200$M total annihilation events per projection. Therefore, all datasets were simulated with experimental conditions at $200$M counts. The {\fontfamily{lmtt}\selectfont DeepCormack} configurations explored in this work are the following:
\begin{enumerate}
     \item \textbf{CNN}: A CNN trained on the projections $f$.
     \item \textbf{MLP}: An MLP trained on the radial density functions $\rho$.
     \item \textbf{UNet}: A UNet trained on the TPMD $\rho^{2\gamma}$.

     \item \textbf{ETE(CNN + MLP + UNet)}: A CNN, MLP, and UNet trained end-to-end.
     
     \item \textbf{CNN$\textsubscript{PT}$ $\rightarrow$ UNet}: Pre-trained CNN, where the outputs are used to train a UNet.
     
     \item \textbf{CNN$\textsubscript{PT}$ + UNet$\textsubscript{PT}$}: Pre-trained CNN and pre-trained UNet linked together and trained.

      \item \textbf{CNN$\textsubscript{PT}$ + MLP$\textsubscript{PT}$ + UNet$\textsubscript{PT}$}: Pre-trained CNN, pre-trained MLP, and pre-trained UNet linked together and trained.

     \item \textbf{CNN$\textsubscript{PT}$ + MLP$\textsubscript{PT}$ $\rightarrow$ UNet}: Pre-trained CNN and pre-trained MLP linked together and trained, then their outputs are used to train a new UNet. A skip connection runs from the CNN through the MCM to recover $\rho^{2\gamma}_P$, and is concatenated as an additional input into the UNet.

\end{enumerate}

\begin{table}[htp!]
\centering
\caption{\footnotesize \textit{Performance of different {\fontfamily{lmtt}\selectfont DeepCormack} model configurations on the models' test data (DMD), and the copper (Cu) and ZrZn$_2$ data with simulated measurement effects applied at $200$M counts. The reconstruction metric is PSNR (mean $\pm$ std), and averaged for different slices between $0.0$ au and $1.1$ au from the centre of the momentum density. The model which give the highest average PSNR for each dataset is given in bold, whilst the second highest is shown with an asterisk.}}
\label{tab:PSNR-Model-Config}
\begin{tabular}{lccc}
\hline
\textbf{Model Configuration}                                & \textbf{DMD}   & \textbf{Cu}      & \textbf{ZrZn$_2$}  \\
\hline
Modified Cormack (Baseline)               & $32.38 \pm 4.14$ & $31.62 \pm 3.00$ & $33.96 \pm 2.08$ \\
CNN                                                         & $33.78 \pm 3.24$ & $32.75 \pm 2.62$ & {${36.20 \pm 1.95}^*$} \\
MLP                                                         & $38.24 \pm 3.52$ & $38.10 \pm 3.44$ & {$30.78 \pm 3.09$} \\ 
UNet                                                        & $38.41 \pm 3.75$ & $\mathbf{40.05 \pm 3.75}$ & {$\mathbf{37.44 \pm 2.57}$} \\
EtE(CNN + MLP + UNet)                                       & ${38.56 \pm 3.69}$ & $36.87 \pm 3.80$ & {$30.00 \pm 2.50$} \\
CNN$_{\text{PT}}$ $\rightarrow$ UNet                        & $37.09 \pm 3.58$ & $37.93 \pm 3.57$ & {$34.94 \pm 2.08$} \\
CNN$_{\text{PT}}$ + UNet$_{\text{PT}}$                      & ${39.15 \pm 3.73}$ & ${39.28 \pm 4.01}^*$ & {$33.51 \pm 2.74$} \\
CNN$_{\text{PT}}$ + MLP$_{\text{PT}}$ + UNet$_{\text{PT}}$  & ${39.10 \pm 3.49}^*$ & $37.62 \pm 3.56$ & {$30.65 \pm 2.74$} \\
CNN$_{\text{PT}}$ + MLP$_{\text{PT}}$ $\rightarrow$ UNet    & $\mathbf{40.69 \pm 3.61}$ & ${38.24 \pm 3.64}$ & {$32.51 \pm 2.92$} \\
\hline
\end{tabular}
\end{table}

Results using Peak Signal to Noise Ratio (PSNR) are show in Table \ref{tab:PSNR-Model-Config}, for all the model configurations and the three test datasets.
The MLP delivers competitive performance on the DMD and Cu datasets ($38.24$ and $38.10$ PSNR respectively), but struggles with ZrZn$_2$, achieving $30.78$ PSNR, far below the MCM baseline of $33.96$ PSNR. On the other hand, the CNN is modest on DMD ($33.78$ PSNR) but achieves $36.20$ PSNR on ZrZn$_2$ — the second highest overall — while retaining the lowest variance of all configurations. The UNet gives the strongest and most consistent results overall, competitive on DMD ($38.41$ PSNR) and the best on both Cu and ZrZn$_2$ ($40.05$ and $37.44$ PSNR respectively). The combination $\text{CNN}_{\text{PT}} + \text{MLP}_{\text{PT}} \rightarrow \text{UNet}$ achieved the highest DMD performance ($40.69$ PSNR), while also achieving $38.24$ PSNR on the copper data. However, this comes at the cost of OOD generalization, with ZrZn$_2$ performance falling to $32.51$ PSNR — below even the Modified Cormack baseline. 

These experiments highlight that wile {\fontfamily{lmtt}\selectfont DeepCormack} has can have far superior performance than baselines, an appropriately designed training dataset is required to fully exploit its potential. For the rest of the experiments, the following configurations are selected and named for convenience: $\text{DC}_\text{C}$ is the CNN only configuration (1), $\text{DC}_\text{M}$ is the MLP only configuration (2), $\text{DC}_\text{U}$ is the UNet only configuration (3) and finally $\text{DC}_\text{CMU}$ is the configuration that performs best in DMD data, i.e. pretraining the CNN and MLP first, and the training all together with the Unet (8).

\begin{figure}[htp!]
  \centering  
  \includegraphics[width=0.95\textwidth]{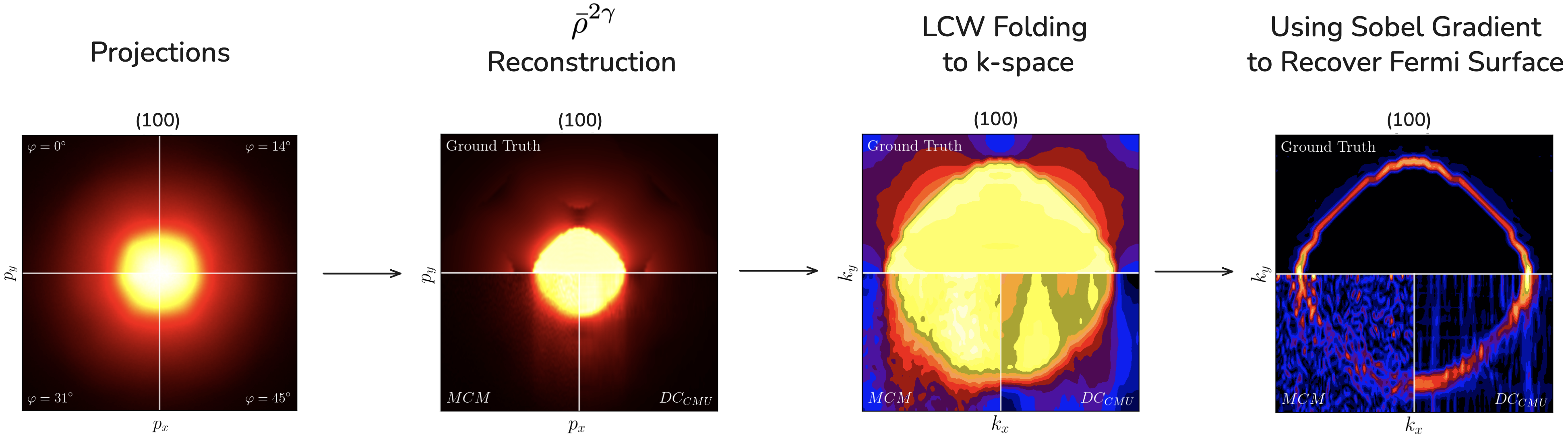}
  \caption{\footnotesize \textit{The steps used for recovering the copper Fermi surface from (i) ACAR projections at $200$M counts, (ii) using the modified Cormack or {\fontfamily{lmtt}\selectfont DeepCormack} methods to recover $\rho^{2\gamma}$ volume, then applying (iii) the LCW theorem to convert the 3D TPMD from \textbf{p}-space to \textbf{k}-space, where (iv) the Fermi surface can be recovered as a discontinuity in the occupation by computing the Sobel gradient.}}
  \label{ResultsFig:Cu_Model_Comparison P-space K-space}
\end{figure}

The goal of ACAR measurements is to recover the Fermi surface, but models output and are trained on TPMDs in \textbf{p}-space. To recover the Fermi surface, the momentum densities for copper were converted to \textbf{k}-space densities using the LCW theorem. Figure \ref{ResultsFig:Cu_Model_Comparison P-space K-space} shows the full pipeline, where projections are converted to TPMDs, \textbf{k}-space momentum densities, before finally extracting the Fermi surface. As this last step — the Fermi surface — is the variable of interest, the rest of the experiments in this work will show results in this space, even though {\fontfamily{lmtt}\selectfont DeepCormack} only reconstructs the first step (from projections to TPMDs).

\subsection{{\fontfamily{lmtt}\selectfont DeepCormack} Performance Across Counts \label{Results: Lowering Counts}}

\subsubsection{Model Performance across counts on DMD Data \label{Results: DMD Reconstruction}}

\begin{figure}[htp!]
  \centering
  \begin{subfigure}[b]{0.45\textwidth}
    \centering
    \includegraphics[width=\textwidth]{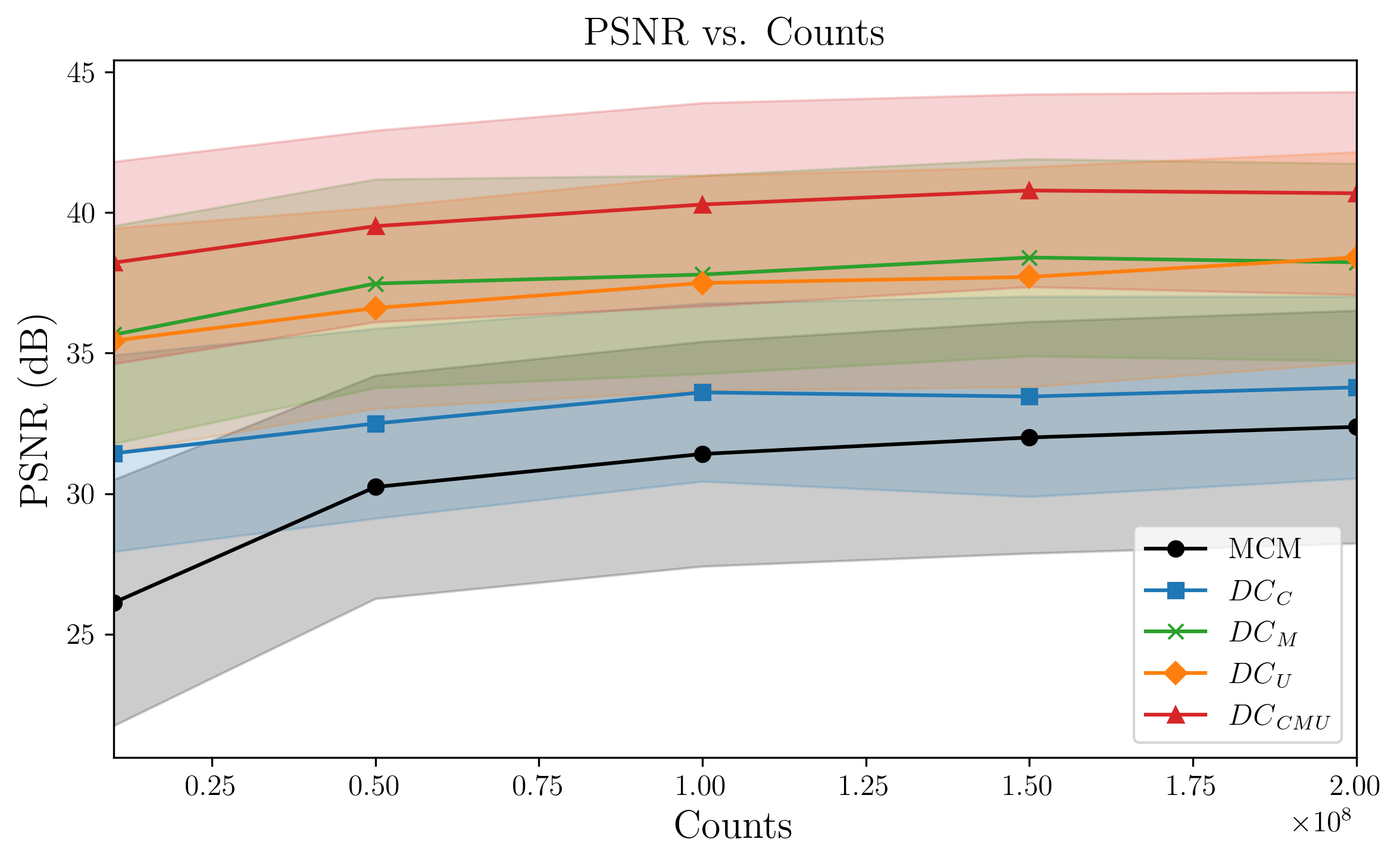}
  \end{subfigure}~
  \begin{subfigure}[b]{0.45\textwidth}
    \centering
    \includegraphics[width=\textwidth]{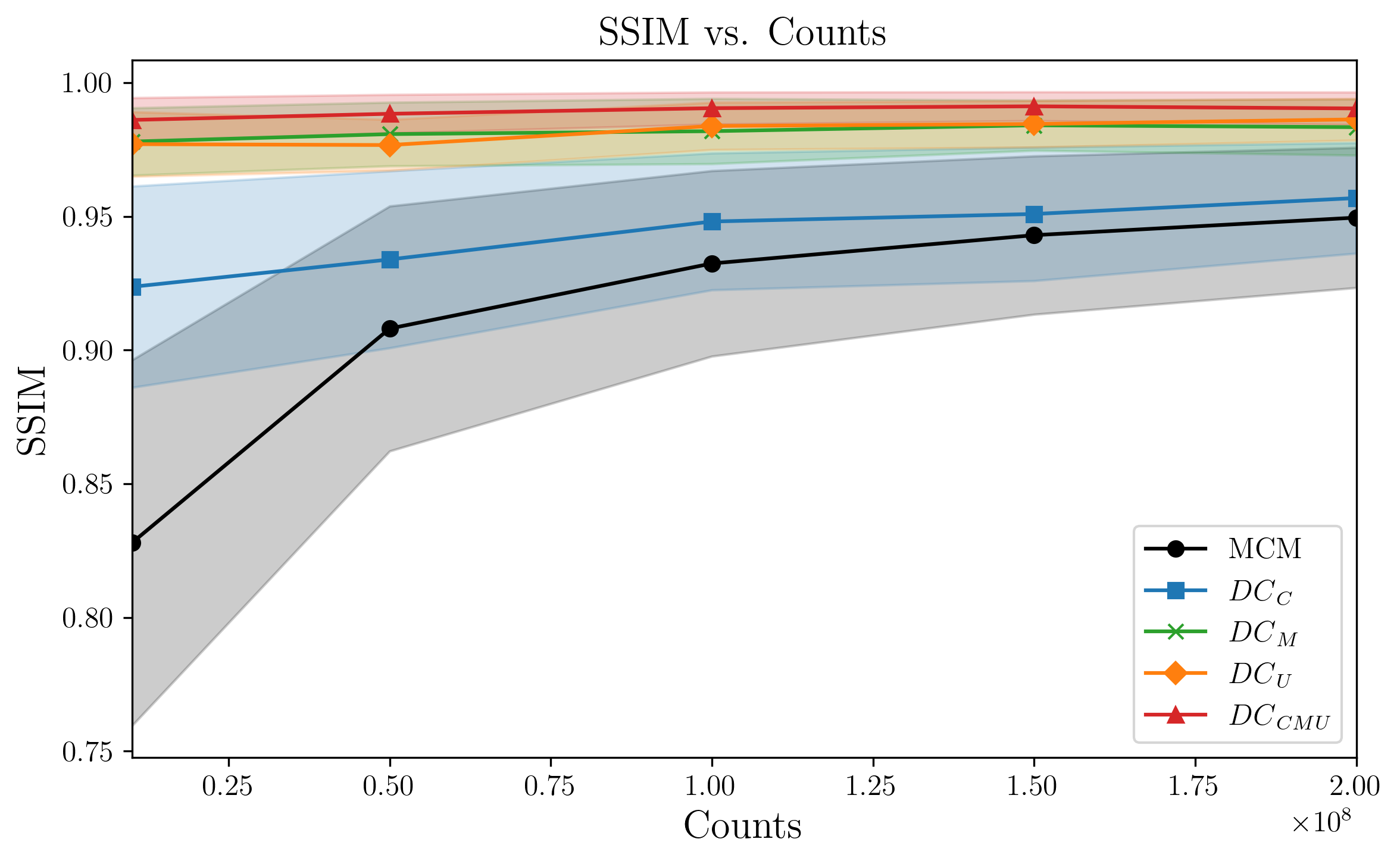}
  \end{subfigure}
  \caption{\footnotesize \textit{The PSNR (i) and SSIM (ii) of the $\text{DC}_\text{C}$, $\text{DC}_\text{M}$, $\text{DC}_\text{U}$, and $\text{DC}_\text{CMU}$ models against the modified Cormack method on the synthetic TPMD data at varying counts.}}
  \label{fig:Experiment2_DMD_LoweringCounts}
\end{figure}

To assess how various {\fontfamily{lmtt}\selectfont DeepCormack}s perform as data quality degrades, the full framework was evaluated on the DMD test set across count levels ranging from $10$M to $200$M, and benchmarked against the MCM using both PSNR and Structural Similarity Index Metric (SSIM) as seen in Table \ref{tab:recon-metrics-counts_DMD} and equivalently in Figure \ref{fig:Experiment2_DMD_LoweringCounts}.

\begin{table}[htp!]
\centering
\caption{\footnotesize \textit{Reconstruction metrics at varying count levels across models  (DMD data).}}
\label{tab:recon-metrics-counts_DMD}
\footnotesize
\begin{tabular}{l|l|ccccc}
\hline
\textbf{Model} & \textbf{Metric} & \textbf{10M} & \textbf{50M} & \textbf{100M} & \textbf{150M} & \textbf{200M} \\
\hline
\multirow{2}{*}{MCM}
 & PSNR (dB)            & $26.12 \pm 4.38$ & $30.23 \pm 3.97$ & $31.41 \pm 4.00$ & $32.00 \pm 4.12$ & $32.38 \pm 4.14$ \\
 & SSIM ($10^{-1}$)     & $8.28 \pm 0.68$  & $9.08 \pm 0.46$  & $9.32 \pm 0.35$  & $9.43 \pm 0.30$  & $9.50 \pm 0.26$  \\
\hline
\multirow{2}{*}{$\text{DC}_\text{C}$}
 & PSNR (dB)            & $31.43 \pm 3.50$ & $32.50 \pm 3.38$ & $33.60 \pm 3.17$ & $33.46 \pm 3.56$ & $33.78 \pm 3.24$ \\
 & SSIM ($10^{-1}$)     & $9.24 \pm 0.38$  & $9.34 \pm 0.33$  & $9.48 \pm 0.26$  & $9.51 \pm 0.25$  & $9.57 \pm 0.21$  \\
\hline
\multirow{2}{*}{$\text{DC}_\text{M}$}
 & PSNR (dB)            & $35.66 \pm 3.88$ & $37.48 \pm 3.71$ & $37.80 \pm 3.53$ & $38.41 \pm 3.51$ & $38.24 \pm 3.52$ \\
 & SSIM ($10^{-1}$)     & $9.78 \pm 0.13$  & $9.81 \pm 0.12$  & $9.82 \pm 0.12$  & $9.84 \pm 0.09$  & $9.83 \pm 0.11$  \\
\hline
\multirow{2}{*}{$\text{DC}_\text{U}$}
 & PSNR (dB)            & $35.44 \pm 4.00$ & $36.61 \pm 3.58$ & $37.50 \pm 3.82$ & $37.71 \pm 3.92$ & $38.41 \pm 3.75$ \\
 & SSIM ($10^{-1}$)     & $9.77 \pm 0.12$  & $9.77 \pm 0.10$  & $9.84 \pm 0.09$  & $9.85 \pm 0.09$  & $9.86 \pm 0.08$  \\
\hline
\multirow{2}{*}{$\text{DC}_\text{CMU}$}
 & PSNR (dB)            & $38.22 \pm 3.60$  & $39.52 \pm 3.41$  & $40.20 \pm 3.62$ & $40.79 \pm 3.43$  & $40.69 \pm 3.61$ \\
 & SSIM ($10^{-1}$)     & $9.86 \pm 0.08$  & $9.88 \pm 0.07$ & $9.90 \pm 0.06$  & $9.91 \pm 0.05$ & $9.90 \pm 0.06$  \\
\hline
\end{tabular}
\end{table}

\FloatBarrier

$\text{DC}_\text{CMU}$ outperforms all models at every count level on both metrics. Most notably, its reconstruction at $10$M counts ($38.22$ PSNR) exceeds the MCM at $200$M counts ($32.38$ PSNR), demonstrating that the framework can maintain reconstruction quality at count levels an order of magnitude lower than those currently used in practice. The MCM degrades with decreasing counts, dropping from $32.38$ to $26.12$ PSNR between $200$M and $10$M, while $\text{DC}_\text{CMU}$ comparatively stable across the same range ($40.69$ to $38.22$ PSNR).

The improvement over the best individual models — the $\text{DC}_{M}$ and $\text{DC}_{U}$, — is more modest, averaging around $+1-1.5$ db PSNR across count levels.

\subsubsection{Model Performance across counts on Copper Data \label{Results: Cu Reconstruction}}

\begin{figure}[htp!]
  \centering
  \begin{subfigure}[b]{0.45\textwidth}
    \centering
    \includegraphics[width=\textwidth]{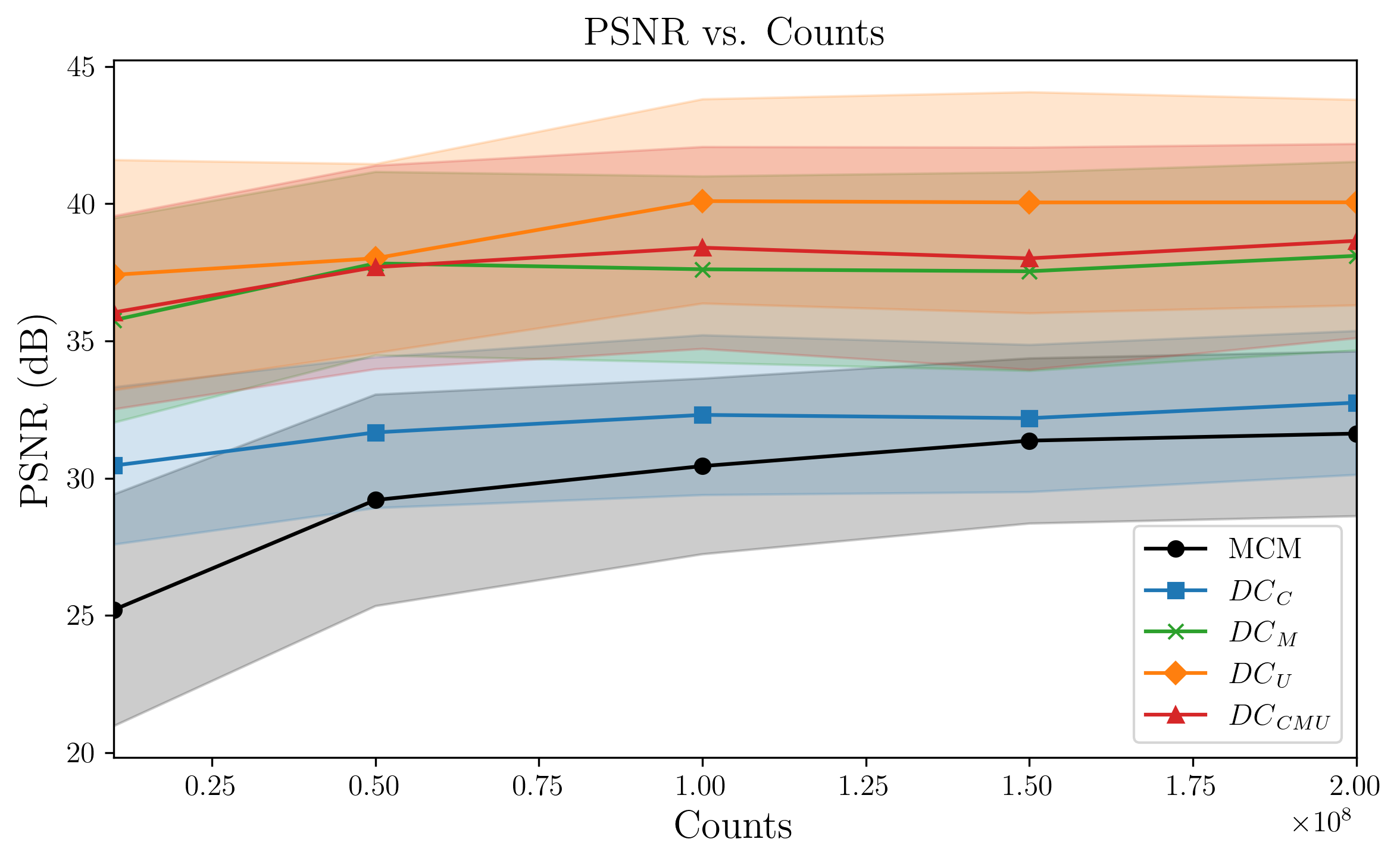}
  \end{subfigure}
  ~
  \begin{subfigure}[b]{0.45\textwidth}
    \centering
    \includegraphics[width=\textwidth]{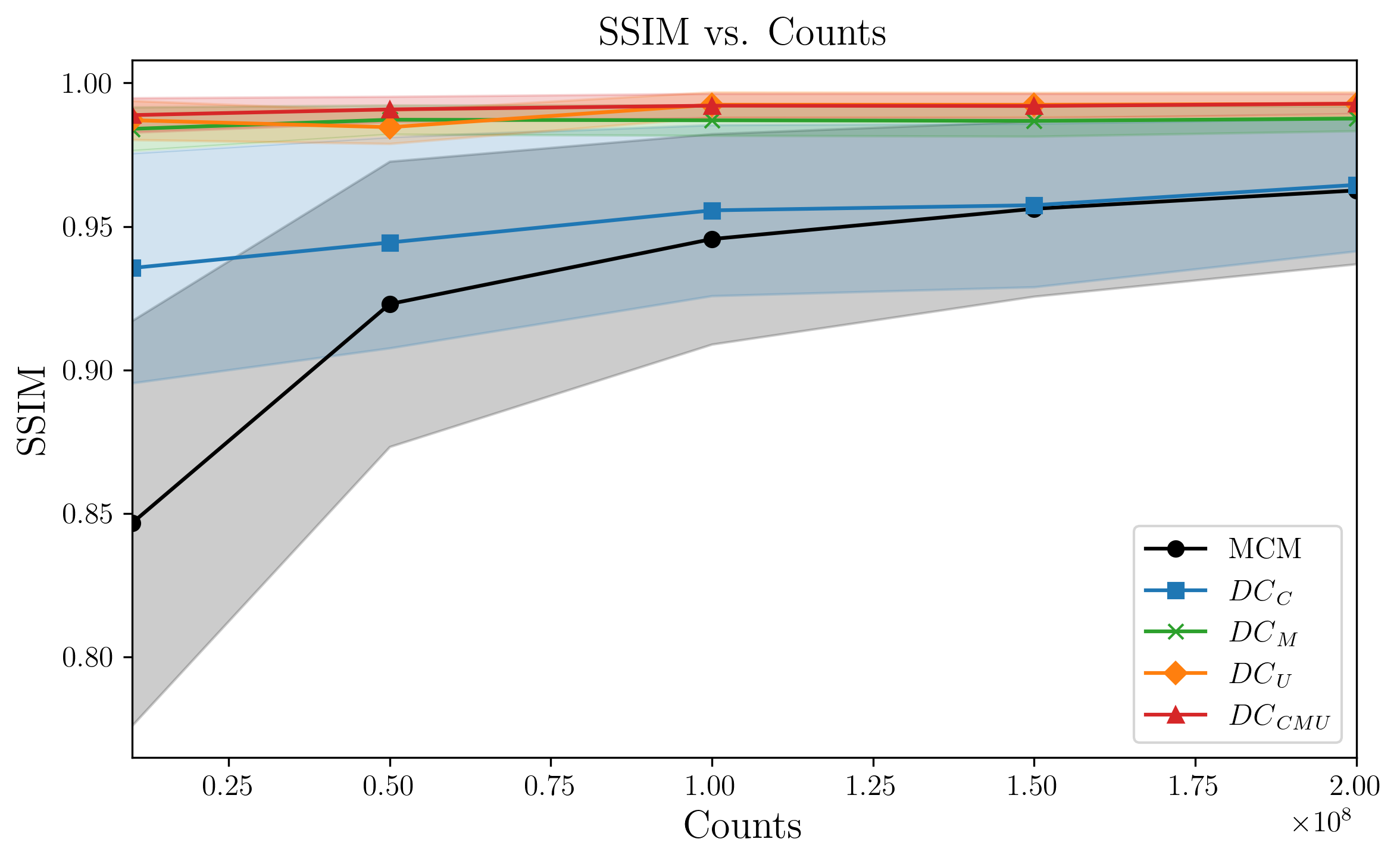}
  \end{subfigure}
  \caption{\footnotesize \textit{The PSNR (i) and SSIM (ii) of the $\text{DC}_\text{C}$, $\text{DC}_\text{M}$, $\text{DC}_\text{U}$, and $\text{DC}_\text{CMU}$ models against the modified Cormack method on the reconstruction from Cu data with simulated measurement conditions.}}
  \label{fig:Experiment2_LoweringCounts}
\end{figure}

Here we test the models on the copper momentum density with simulated measurement conditions applied for counts ranging from $10-200$M. Results can be seen in Table \ref{tab:recon-metrics-counts-cu} ans Figure \ref{fig:Experiment2_LoweringCounts}. This is the same reference used to generate synthetic training data. Since the copper features are not far to the training distribution, this will test how well the models generalize to real TPMD data rather than purely synthetically generated ones.

\begin{table}[htp!]
\centering
\caption{\footnotesize \textit{Reconstruction metrics at varying count levels across models (Cu data).}}
\label{tab:recon-metrics-counts-cu}
\footnotesize
\begin{tabular}{l|l|ccccc}
\hline
\textbf{Model} & \textbf{Metric} & \textbf{10M} & \textbf{50M} & \textbf{100M} & \textbf{150M} & \textbf{200M} \\
\hline
\multirow{2}{*}{MCM}
 & PSNR (dB)            & $25.20 \pm 4.22$ & $29.20 \pm 3.85$ & $30.44 \pm 3.20$ & $31.37 \pm 3.01$ & $31.62 \pm 3.00$ \\
 & SSIM ($10^{-1}$)     & $8.47 \pm 0.71$  & $9.23 \pm 0.50$  & $9.46 \pm 0.37$  & $9.56 \pm 0.31$  & $9.63 \pm 0.26$  \\
\hline
\multirow{2}{*}{$\text{DC}_\text{C}$}
 & PSNR (dB)            & $30.46 \pm 2.87$ & $31.67 \pm 2.75$ & $32.31 \pm 2.91$ & $32.19 \pm 2.68$ & $32.75 \pm 2.62$ \\
 & SSIM ($10^{-1}$)     & $9.36 \pm 0.40$  & $9.44 \pm 0.37$  & $9.56 \pm 0.30$  & $9.57 \pm 0.29$  & $9.65 \pm 0.23$  \\
\hline
\multirow{2}{*}{$\text{DC}_\text{M}$}
 & PSNR (dB)            & $35.76 \pm 3.73$ & $37.83 \pm 3.35$ & $37.61 \pm 3.40$ & $37.54 \pm 3.62$ & $38.10 \pm 3.44$ \\
 & SSIM ($10^{-1}$)     & $9.84 \pm 0.08$  & $9.87 \pm 0.05$  & $9.87 \pm 0.05$  & $9.87 \pm 0.05$  & $9.88 \pm 0.04$  \\
\hline
\multirow{2}{*}{$\text{DC}_\text{U}$}
 & PSNR (dB)            & $37.41 \pm 4.20$ & $38.01 \pm 3.44$ & $40.10 \pm 3.72$ & $40.05 \pm 4.03$ & $40.05 \pm 3.75$ \\
 & SSIM ($10^{-1}$)     & $9.87 \pm 0.07$  & $9.85 \pm 0.06$  & $9.92 \pm 0.04$  & $9.92 \pm 0.04$  & $9.93 \pm 0.04$  \\
\hline
\multirow{2}{*}{$\text{DC}_\text{CMU}$}
 & PSNR (dB)            & $36.04 \pm 3.52$ & $37.69 \pm 3.71$ & $38.24 \pm 3.64$ & $38.01 \pm 4.05$ & $38.65 \pm 3.54$ \\
 & SSIM ($10^{-1}$)     & $9.89 \pm 0.06$ & $9.91 \pm 0.05$ & $9.92 \pm 0.04$  & $9.92 \pm 0.04$ & $9.93 \pm 0.04$  \\
\hline

\end{tabular}
\end{table}

\FloatBarrier

On the Cu data, the $\text{DC}_\text{U}$ is the strongest performer across all count levels, reaching $40.05$ PSNR at $200$M counts and $37.41$ at $10$M, outperforming $\text{DC}_\text{CMU}$, which achieves $38.65$ and $36.04$ dB respectively. $\text{DC}_\text{CMU}$ performs comparably to the $\text{DC}_\text{M}$ across all counts, around $1.40$ PSNR lower than the $\text{DC}_\text{U}$. However, all models again comfortably outperform the MCM at every count level. Notably, $\text{DC}_\text{CMU}$ at $10$M counts ($36.04$ PSNR) still exceeds the MCM at $200$M counts ($31.62$ dB PSNR), consistent with the DMD results.

When these slices are individually passed through {\fontfamily{lmtt}\selectfont DeepCormack}, they need to be renormalized in order to produce coherent 3D TPMDs and Fermi surfaces. This is done using a smoothed version of the measured momentum density data's maximum intensity across slices.

Despite the PSNR and SSIM gap between the $\text{DC}_\text{U}$ and $\text{DC}_\text{CMU}$ in momentum space, the Fermi surface reconstructions show that both produce qualitatively comparable results across the full count range, recovering a greater detail of features along all three slices than the MCM. Notably, the recovered Fermi surface remains largely intact even at $10$M counts, with features that are more distinct than the MCM at $200$M counts.

\begin{figure}[htp!]
  \centering  
  \includegraphics[width=0.95\textwidth]{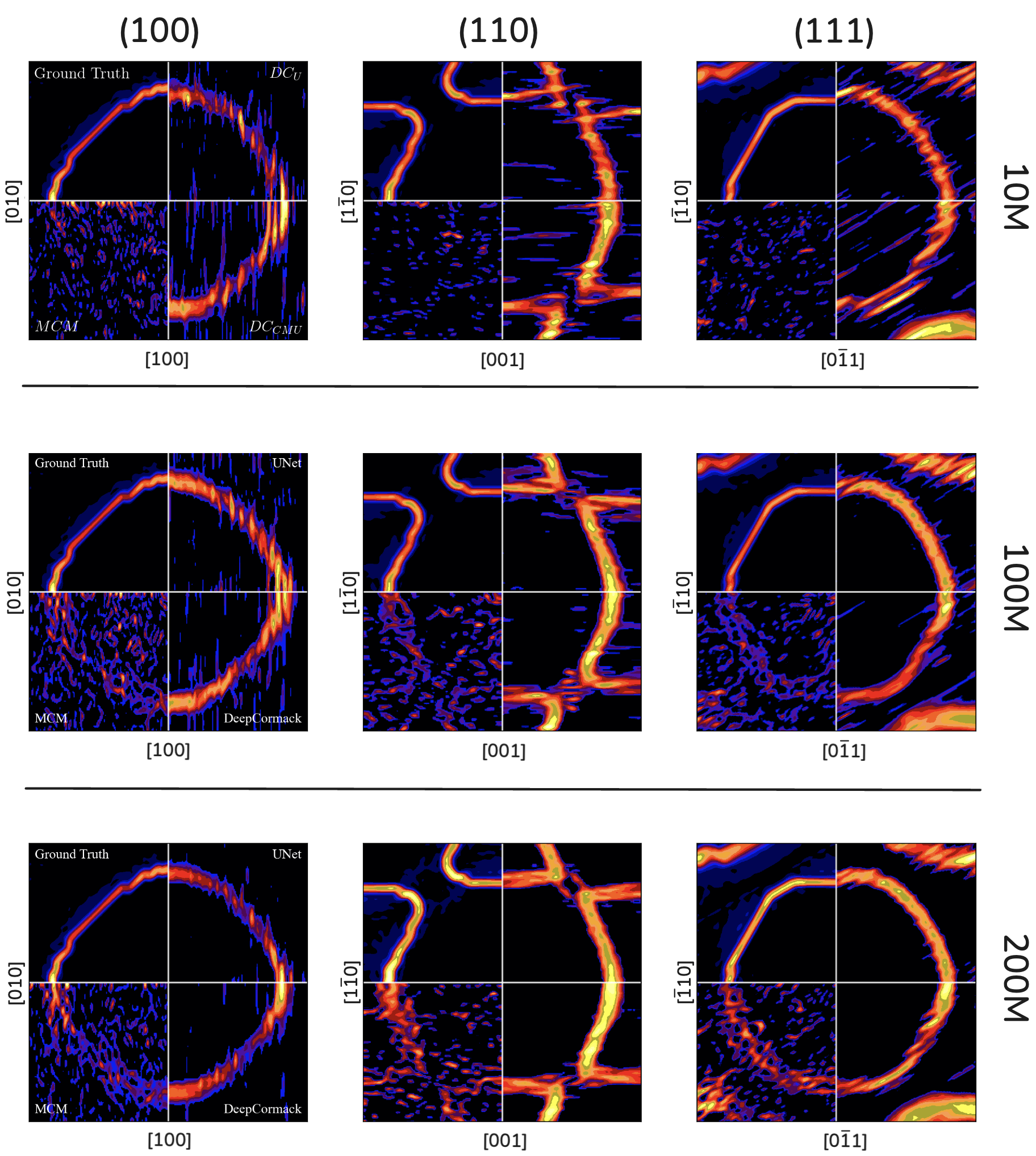}
  \caption{\footnotesize \textit{The Cu Fermi surface as reconstructed by the $\text{DC}_\text{U}$, $\text{DC}_\text{CMU}$, and MCM compared against the ground truth at (from top to bottom) $200$M, $100$M, and $10$M counts under simulated measurement conditions. Columns showcase slices through different axis of the 3D Fermi surface.}}
  \label{ResultsFig:Cu_Model_Comparison}
\end{figure}

\FloatBarrier

\subsubsection{Model Performance on ZrZn\texorpdfstring{$_2$}{2} Data \label{Results: ZrZn2 Reconstruction}}

Here the models are tested on ZrZn$_2$ data with simulated measurement conditions applied across $10$M to $200$M counts (Table \ref{tab:recon-metrics-counts-zrzn2}, Figure \ref{fig:Experiment2_LoweringCounts}). Unlike the copper data, the ZrZn$_2$ momentum density has features that are outside the training distribution, making this a strict test of generalization.

\begin{table}[htp!]
\centering
\caption{\footnotesize \textit{Reconstruction metrics at varying count levels across models (ZrZn$_2$ data).}}
\label{tab:recon-metrics-counts-zrzn2}
\footnotesize
\begin{tabular}{l|l|ccccc}
\hline
\textbf{Model} & \textbf{Metric} & \textbf{10M} & \textbf{50M} & \textbf{100M} & \textbf{150M} & \textbf{200M} \\
\hline
\multirow{2}{*}{MCM}
 & PSNR (dB)            & $26.13 \pm 2.43$ & $30.64 \pm 2.63$ & $32.39 \pm 2.19$ & $33.38 \pm 2.35$ & $33.96 \pm 2.08$ \\
 & SSIM ($10^{-1}$)     & $8.97 \pm 0.38$  & $9.51 \pm 0.24$  & $9.65 \pm 0.18$  & $9.72 \pm 0.14$  & $9.76 \pm 0.11$  \\
\hline
\multirow{2}{*}{$\text{DC}_\text{C}$}
 & PSNR (dB)            & $33.98 \pm 2.38$ & $35.17 \pm 1.94$ & $35.79 \pm 1.98$ & $35.77 \pm 1.99$ & $36.20 \pm 1.95$ \\
 & SSIM ($10^{-1}$)     & $9.65 \pm 0.16$  & $9.73 \pm 0.12$  & $9.77 \pm 0.09$  & $9.77 \pm 0.08$  & $9.80 \pm 0.07$  \\
\hline
\multirow{2}{*}{$\text{DC}_\text{M}$}
 & PSNR (dB)            & $30.65 \pm 2.50$ & $30.71 \pm 2.81$ & $30.69 \pm 2.69$ & $30.73 \pm 2.67$ & $30.78 \pm 3.09$ \\
 & SSIM ($10^{-1}$)     & $8.90 \pm 0.24$  & $8.95 \pm 0.27$  & $8.91 \pm 0.31$  & $8.94 \pm 0.33$  & $8.94 \pm 0.32$  \\
\hline
\multirow{2}{*}{$\text{DC}_\text{U}$}
 & PSNR (dB)            & $35.02 \pm 1.94$ & $36.60 \pm 2.68$ & $36.87 \pm 2.04$ & $37.49 \pm 2.48$ & $37.44 \pm 2.57$ \\
 & SSIM ($10^{-1}$)     & $9.76 \pm 0.09$  & $9.66 \pm 0.19$  & $9.79 \pm 0.15$  & $9.85 \pm 0.05$  & $9.85 \pm 0.04$  \\
\hline
\multirow{2}{*}{$\text{DC}_\text{CMU}$}
 & PSNR (dB)            & $32.32 \pm 2.59$ & $32.47 \pm 3.17$ & $32.51 \pm 2.92$ & $32.83 \pm 2.39$ & $32.82 \pm 2.46$ \\
 & SSIM ($10^{-1}$)     & $9.31 \pm 0.19$ & $9.20 \pm 0.23$ & $9.36 \pm 0.26$  & $9.25 \pm 0.29$ & $9.28 \pm 0.31$  \\
\hline
\end{tabular}
\end{table}

\begin{figure}[htp!]
  \centering
  \begin{subfigure}[b]{0.45\textwidth}
    \centering
    \includegraphics[width=\textwidth]{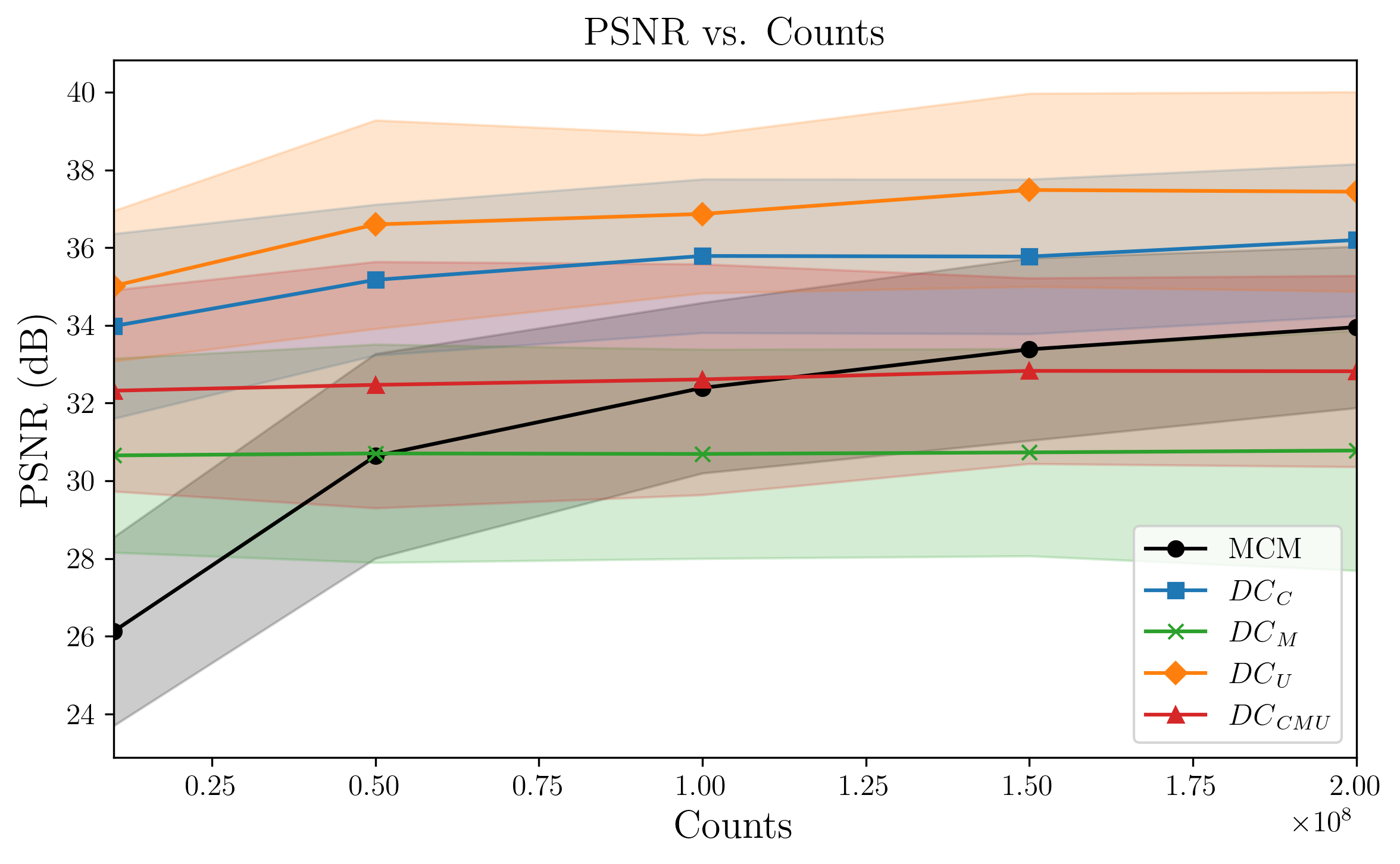}
  \end{subfigure}
  ~
  \begin{subfigure}[b]{0.45\textwidth}
    \centering
    \includegraphics[width=\textwidth]{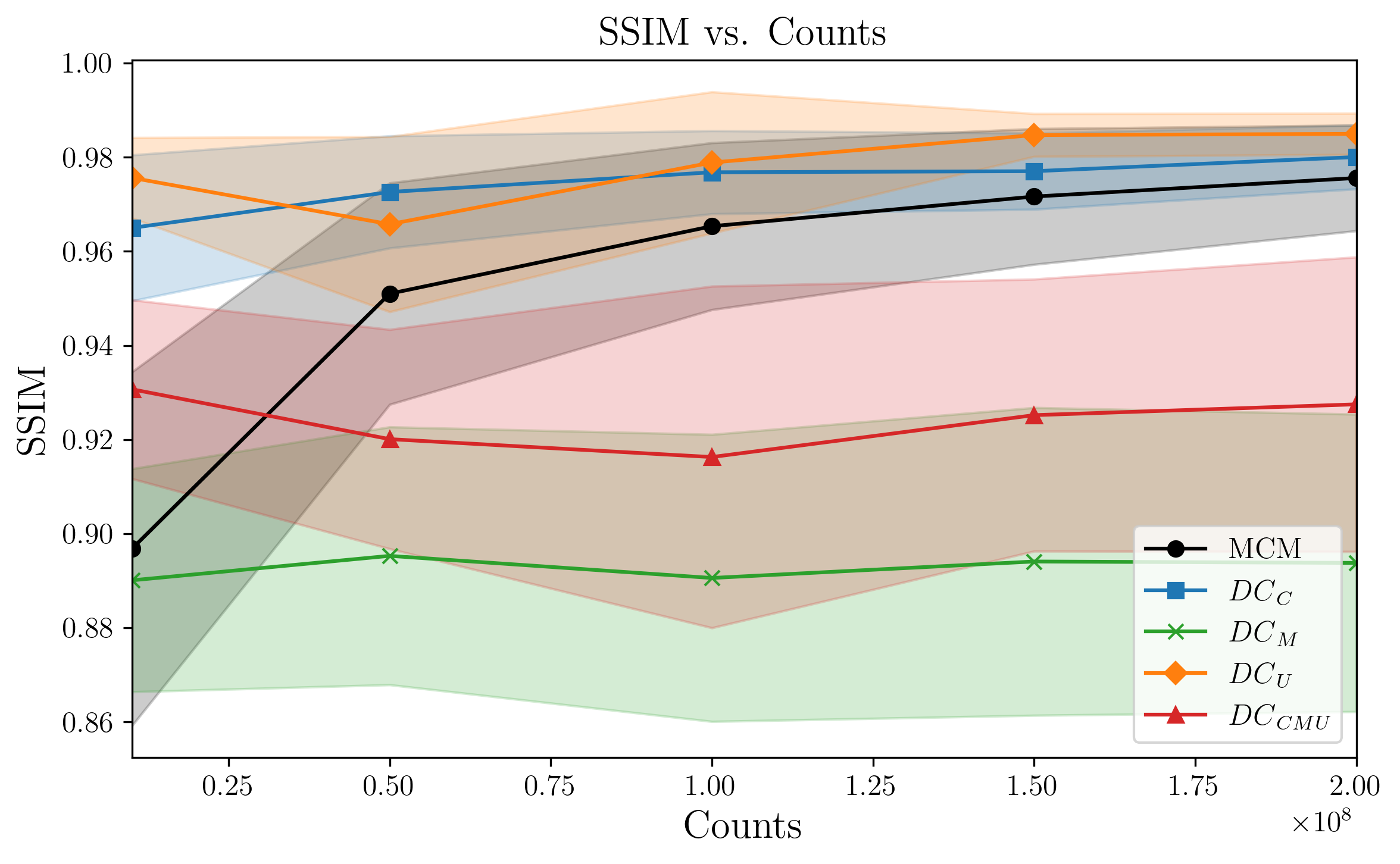}
  \end{subfigure}
  \caption{\footnotesize \textit{The PSNR (i) and SSIM (ii) of the $\text{DC}_\text{C}$, $\text{DC}_\text{M}$, $\text{DC}_\text{U}$, and $\text{DC}_\text{CMU}$ models against the modified Cormack method on the reconstruction from ZrZn$_2$ data.}}
  
  \label{fig:Experiment2_ZrZn2_LoweringCounts}
\end{figure}
\FloatBarrier

\begin{figure}[htp!]
  \centering  
  \includegraphics[width=0.95\textwidth]{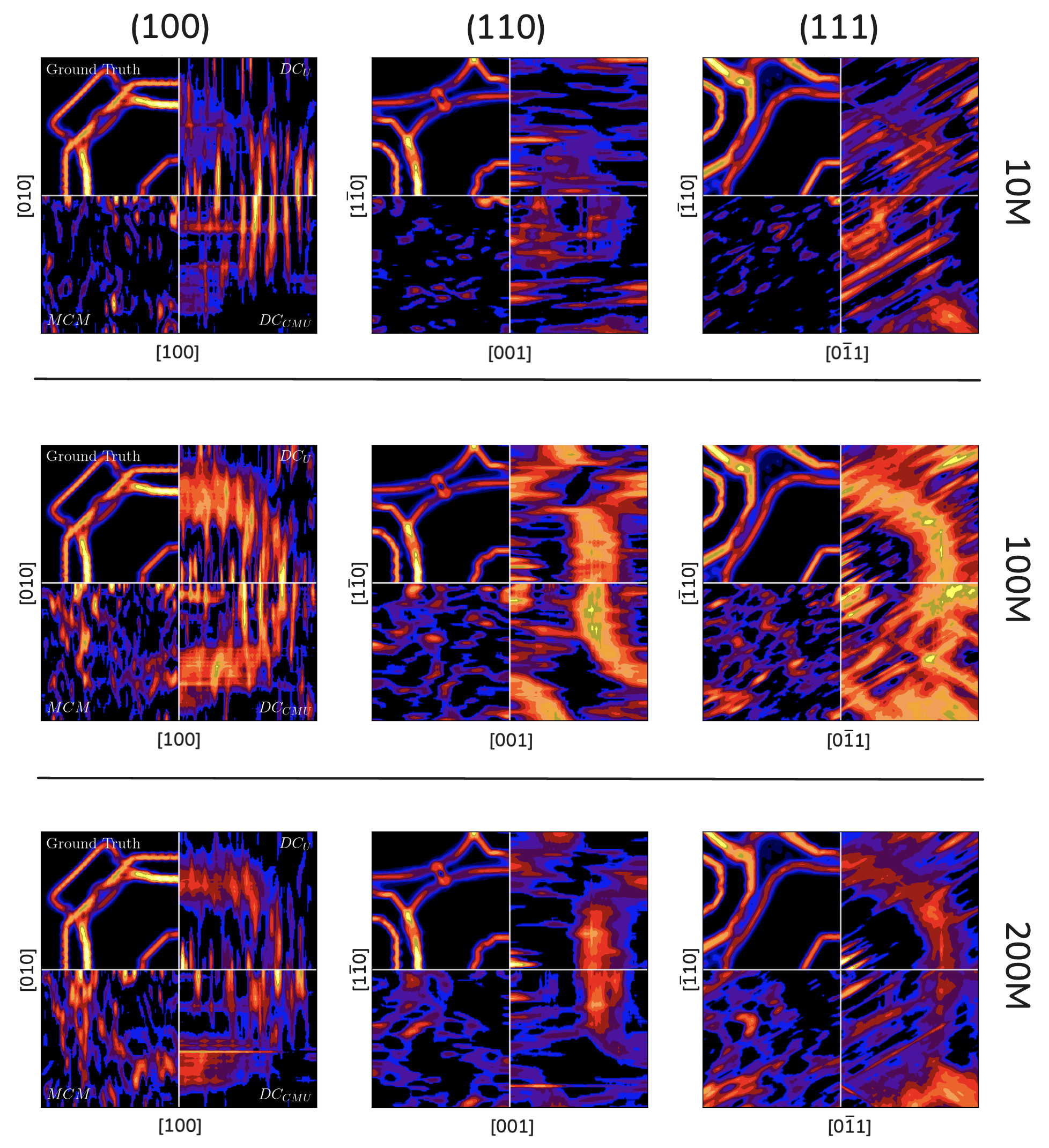}

  \caption{\footnotesize \textit{The ZrZn$_2$ Fermi surface as reconstructed by the $\text{DC}_\text{U}$, $\text{DC}_\text{CMU}$, and MCM compared against the ground truth at $200$M, $100$M, and $10$M counts under simulated measurement conditions.}}
  \label{ResultsFig:ZrZn2_Model_Comparison}
\end{figure}

$\text{DC}_\text{CMU}$s performance is significantly lower on the ZrZn$_2$ dataset, achieving only $32.82$ PSNR at $200$M counts, which is below the MCM ($33.96$ PSNR) and the $\text{DC}_\text{C}$ ($36.20$ PSNR), but greater than that of the $\text{DC}_\text{M}$ by itself ($30.78$ PSNR). This performance drop becomes apparent when computing and comparing the Fermi surfaces, as shown in Figure \ref{ResultsFig:ZrZn2_Model_Comparison}.

\begin{figure}[htp!]
  \centering  
  \includegraphics[width=0.95\textwidth]{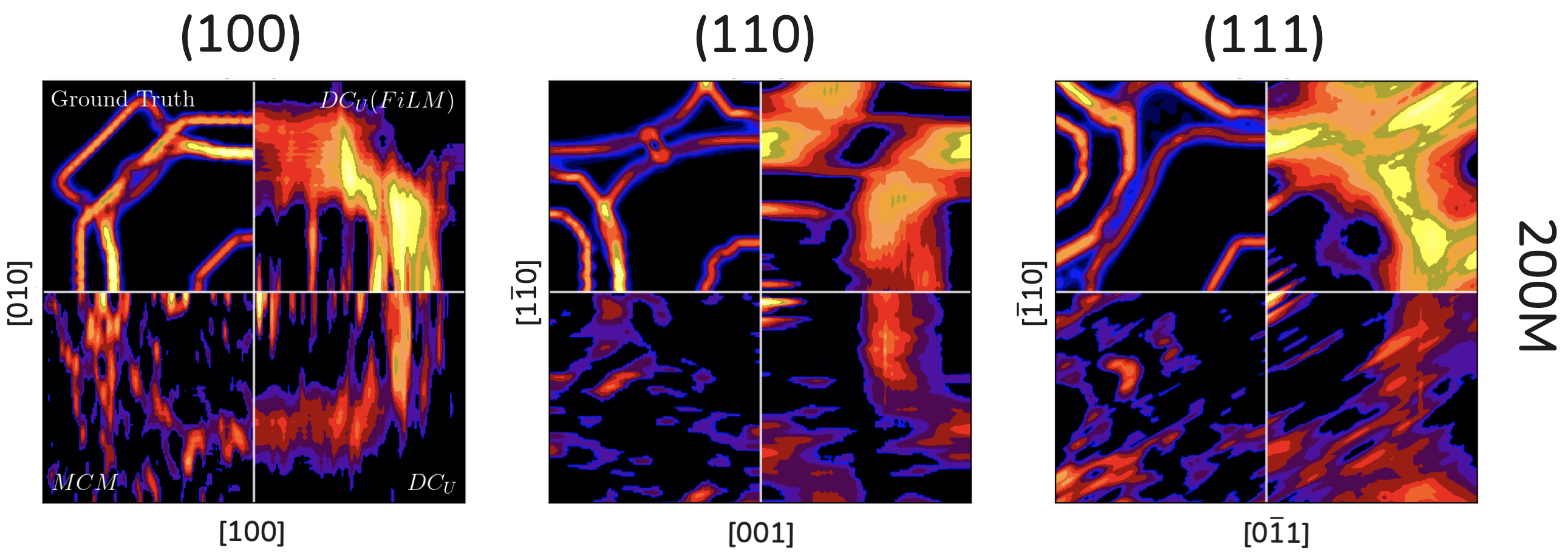}

  \caption{\footnotesize \textit{Comparison of the ZrZn$_2$ Fermi surface reconstruction for $\text{DC}_{U}$ with and without FiLM conditioning on the noise. The ZrZn$_2$ simulated measurement data was passed through the models at $200$M counts}}
  \label{ResultsFig: MCM+DC+UNet vs UNet FiLM ZrZn2 Experimental}
\end{figure}

We found out, however, that conditioning the $\text{DC}_\text{U}$ with FiLM conditioning has a noticeable impact on the reconstructed Fermi surface for ZrZn$_2$, as seen in Figure \ref{ResultsFig: MCM+DC+UNet vs UNet FiLM ZrZn2 Experimental}. Qualitatively, the results suggest that $\text{DC}_\text{U}$ produces a better Fermi surface than MCM, albeit its unclear if a sufficiently informative one. 

\FloatBarrier

\section{Discussion \label{Discussion}}

In this work, we proposed a {\fontfamily{lmtt}\selectfont DeepCormack} family of models for Fermi surface tomography, together with data generation approaches and testing on a variety of count levels, which is directly proportional to scanning time.

A large combination of different {\fontfamily{lmtt}\selectfont DeepCormack} architecture variations were tested, and results were generally inconclusive. Each configuration carries its own trade-offs. The CNN and MLP are the least intensive to train: the CNN generalizes well but gives weaker in-distribution performance, while the MLP is powerful within the training distribution but struggles on OOD data. The UNet is the most computationally demanding to train, but offers the strongest and most reliable performance overall. We note, however, that the UNet's advantage likely reflects the limitations of the training data as much as its architectural strengths. There are currently no public databases of experimentally measured or DFT-generated momentum densities for real samples that could support supervised learning, so models were trained entirely on synthetic data derived from a copper reference. With more diverse and representative training data, configurations that perform strongest on the DMD test set — particularly $\text{DC}_{\text{CMU}}$, which achieves amongst the highest in-distribution performance while offering considerable architectural flexibility — may prove more competitive. Each component is straightforward to train independently and can subsequently be linked and fine-tuned for a small number of epochs, making this framework practical and adaptable, particularly as richer training data becomes available. Furthermore, if strong reconstruction performance can be maintained at lower count levels, collecting data along more than $5$ projection directions becomes viable, and the richer angular sampling may allow the full range of model configurations to reach their potential.

Notably, we saw that the performance gain of the models depend on which data are tested. While pre-training can improve reconstruction quality on data close to the training distribution, it appears to specialize the models towards training-data-like features in a way that does not fully recover with joint training. More broadly, models with sufficient capacity to exploit the structure of the synthetic data tend to do so at the cost of generalization, and the  $\text{DC}_\text{U}$ currently strikes the best balance between the two. It should be noted that these results are obtained on synthetic DMD data drawn from the same distribution as the training data, which naturally favours all models evaluated here. Real experimental data will differ in both noise characteristics and momentum density features, and how well these results carry over depends on how faithfully the simulated data reproduces real measurements — as assessed in Section \ref{Results: Validating Sim Approach}. Training on a broader range of DFT-generated momentum densities, or ideally real experimental data, would likely improve performance and generalization across the board. Testing the models on real experimental data will be explored in future work.

That said, the lack of generalization to out-of-distribution samples is a limitation, but only a moderate one. As is true of inverse problems in general, learned reconstructions are highly sensitive to the data distribution, meaning they tend to generalize only moderately, particularly when tested on real data. An advantage here is that Fermi surface tomography is a very slow process — one that can rely on precomputed approximations of the ground truth. Currently, ACAR measurements for $200$M counts take over $3$ months of scanning time. However, one can pre-compute a reference momentum density within a day or so. This reference can then be used to generate training data for {\fontfamily{lmtt}\selectfont DeepCormack} via DMD, after which one can choose to reduce the measurement time. Our work suggests that this will still produce higher quality Fermi surfaces than the modified Cormack method, in significantly less time. That said, {\fontfamily{lmtt}\selectfont DeepCormack} produces significantly better reconstructions than MCM on $200$M-count data, so even without the time-reduction possibility, {\fontfamily{lmtt}\selectfont DeepCormack} enables much higher quality Fermiology via 2D-ACAR. It is also noteworthy that FiLM conditioning led to some qualitatively better reconstructions than MCM on OOD experimental data, meaning OOD testing cannot be completely discarded.

In general, for this work it is worth noting that we used the natural image quality metrics PSNR and SSIM to evaluate performance. While, in broad terms, the qualitative evaluations of the results here correlate with an increase in PSNR and SSIM, these metrics may not be the appropriate ones for evaluating fine differences or for choosing a ''best" configuration of {\fontfamily{lmtt}\selectfont DeepCormack}. Ideally, the quality of the Fermi surface — or, better still, the quality of quantitative values extracted from the Fermi surface — would be the best metric, as these are what physicists will use, not the images themselves. This is, however, non-trivial, and would require a set of labelled ground truth data, which is not available for this task at the time of writing. Future work should focus on validating {\fontfamily{lmtt}\selectfont DeepCormack} with such metrics, evaluating how useful it is for real outcomes with real data. This idea could also be extended to training the models. If MSE/PSNR of the TPMDs that {\fontfamily{lmtt}\selectfont DeepCormack} outputs is not the goal, moving the training loss to metrics defined on the Fermi surface would be an improved solution. This is non-trivial, however, as applying the LCW theorem to pseudo-real data is not obvious, and computationally expensive, as it would require storing all TPMD slices during training rather than just a subset.

All configurations here have been desiged to only $5$ projections, which constrains the quality of reconstruction regardless of the model used. Given the strong performance of {\fontfamily{lmtt}\selectfont DeepCormack}, a natural next step would be to acquire more projections — for example 10-20 — at proportionally lower count levels ($50$ — $25$\% of $200$M counts), keeping total acquisition time fixed. {\fontfamily{lmtt}\selectfont DeepCormack} could compensate for the lower quality projections, whilst the richer angular sampling could improve reconstruction quality overall. This may also benefit configurations beyond the standalone $\text{DC}_\text{U}$, as the CNN and MLP are well-suited to exploiting additional input channels, and modular frameworks such as $\text{DC}_\text{CMU}$ could prove better positioned to take advantage of both denser projection coverage and richer training data.

\section{Conclusions \label{Conclusions}}

In this work we proposed {\fontfamily{lmtt}\selectfont DeepCormack}, a family of data-driven, model-based reconstruction algorithms for Fermi surface tomography. We also propose a data generation model based on Koopman operators to train such models. Results show significant improvement in reconstruction quality compared to the baseline, standard Modified Cormack Method, both in the standard acquisition regime and in significantly reduced ones. This not only allows for much higher quality results for the condensed matter physics community, but also potentially saves weeks of scanning time without any loss (in fact, with an increase) in reconstruction quality.

\newpage

\bibliographystyle{unsrt}
\bibliography{sample}

\newpage

\appendix
\section{The TPMD Reconstruction Step in the MCM \label{Appendix:MCM_Matrix}}
\begin{figure}[htp!]
    \centering
    \includegraphics[width=0.95\textwidth]{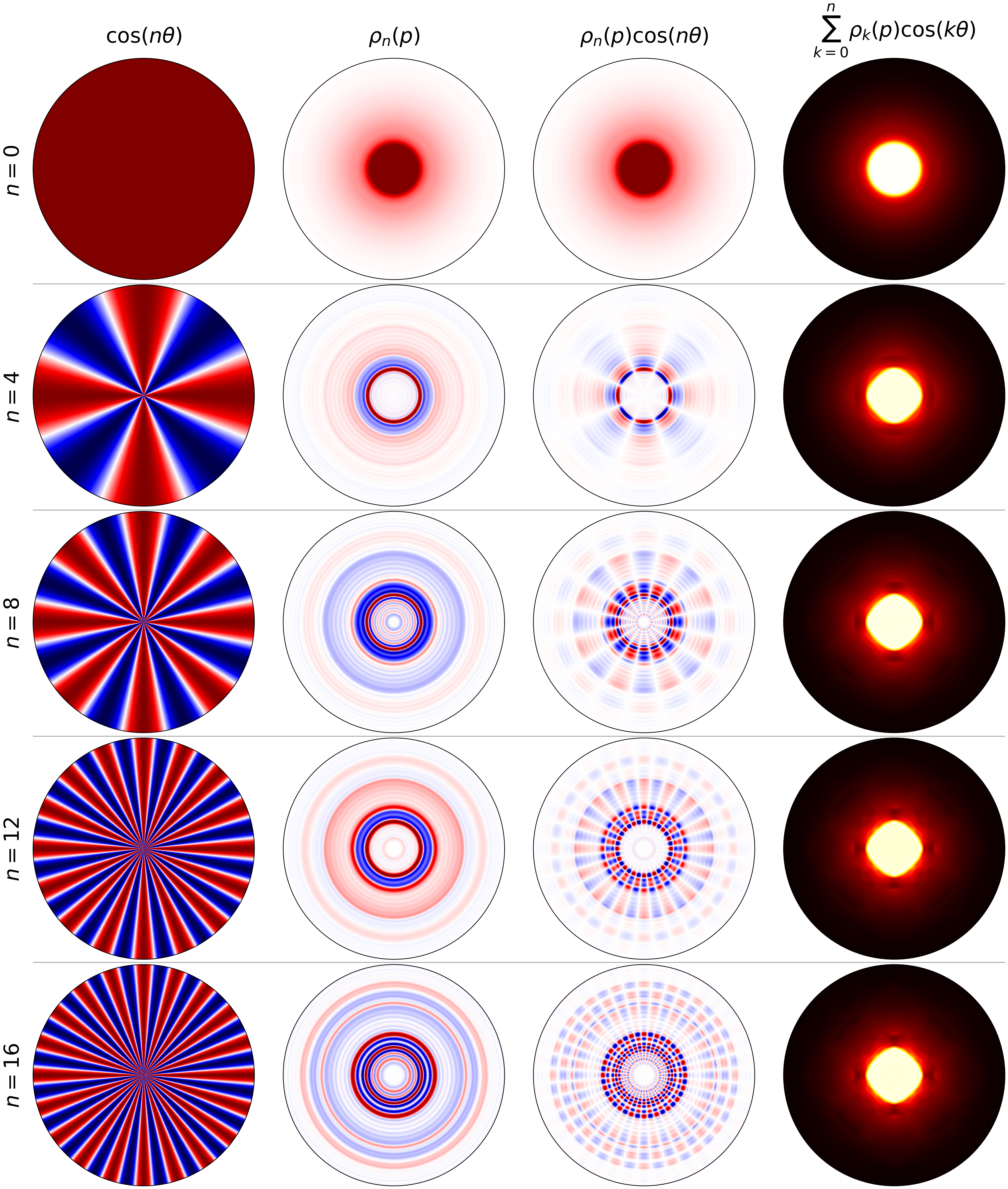} 
    \caption{\footnotesize{\textit{The final step of the MCM reconstruction process for copper, which has four-fold symmetry. The radial density functions $\rho_n(p)$ (here shown normalized) are multiplied by their corresponding cosine modulation term $\cos(n\theta)$ around a unit circle. The final column shows the effect of successively adding together higher order terms to recover the central-slice TPMD for copper — with no noise and smearing, this converges towards the ground truth.}}}
    \label{AppendixFig:Final_MCM_Step}
\end{figure}
\FloatBarrier

\newpage
\section{Dynamic Mode Decomposition for 3D TPMD Generation \label{Appendix:DMD_3D_TPMD}}

\begin{figure}[htp!]
  \centering
  \includegraphics[width=0.98\textwidth]{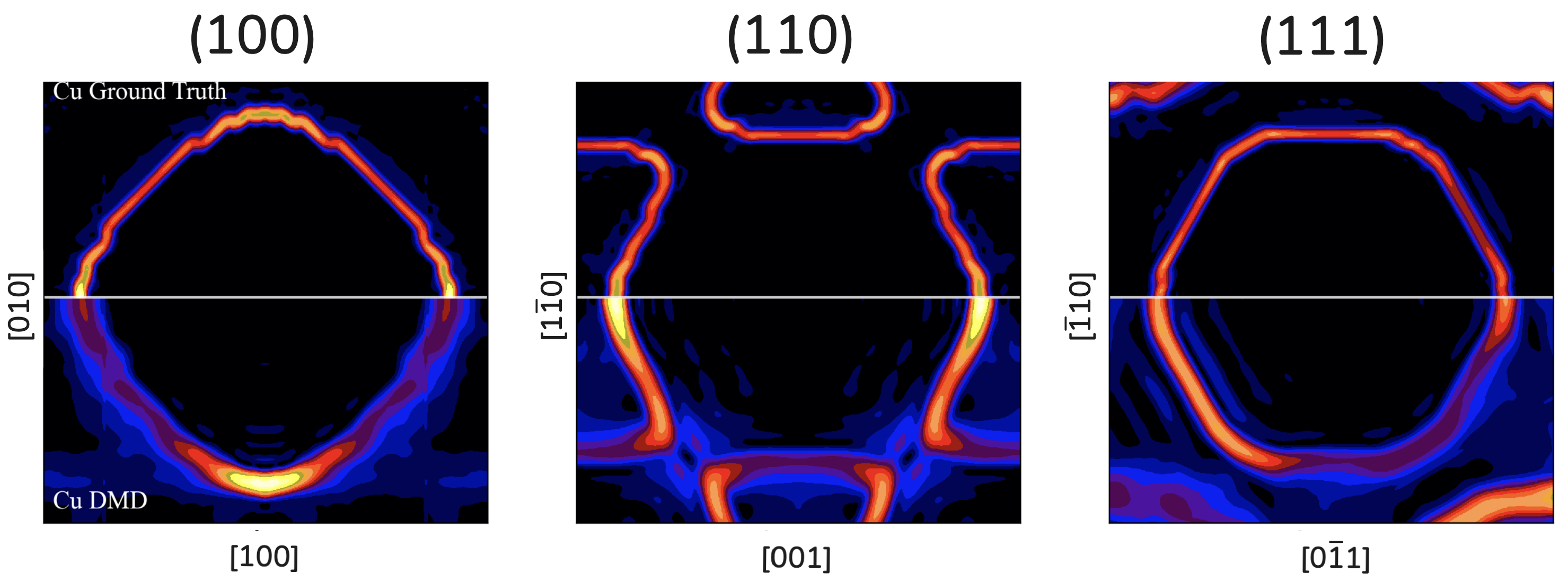}
  \caption{\footnotesize{\textit{The difference between the ground truth Fermi surface for copper and its DMD evolved counterpart: taking the Koopman operators computed from the GT copper momentum density to evolve its GT central slice into a 3D TPMD. This is shown through the high-symmetry planes $(100)$, $(110)$, and $(111)$, illustrating the information lost across the Fermi surface. }}}
  \label{AppendixFig:DMD_for_Data_Generation}
\end{figure}

\paragraph{Copper Koopman Operator on Copper Central Slice:} The copper momentum density is where we derived our per-channel Koopman operators $\mathcal{K}_n$ from. Here, we evaluate whether evolving the Chebyshev coefficients for the copper central slice via dynamic mode decomposition is sufficient to recover the original structure. To do this comparison, we took the original DFT-generated copper momentum density along with the DMD-evolved copper momentum density, applied LCW-folding to convert from $p$-space to $k$-space before extracting their Fermi surfaces. Figure \ref{AppendixFig:DMD_for_Data_Generation} compares slices taken through high symmetry planes $(100)$, $(110)$, and $(111)$ of the Fermi surfaces.

\newpage
\section{Synthetic 3D TPMDs Examples \label{Appendix:SynthDataGen}}

\begin{figure}[htp!]
  \centering  
  \includegraphics[width=0.98\textwidth]{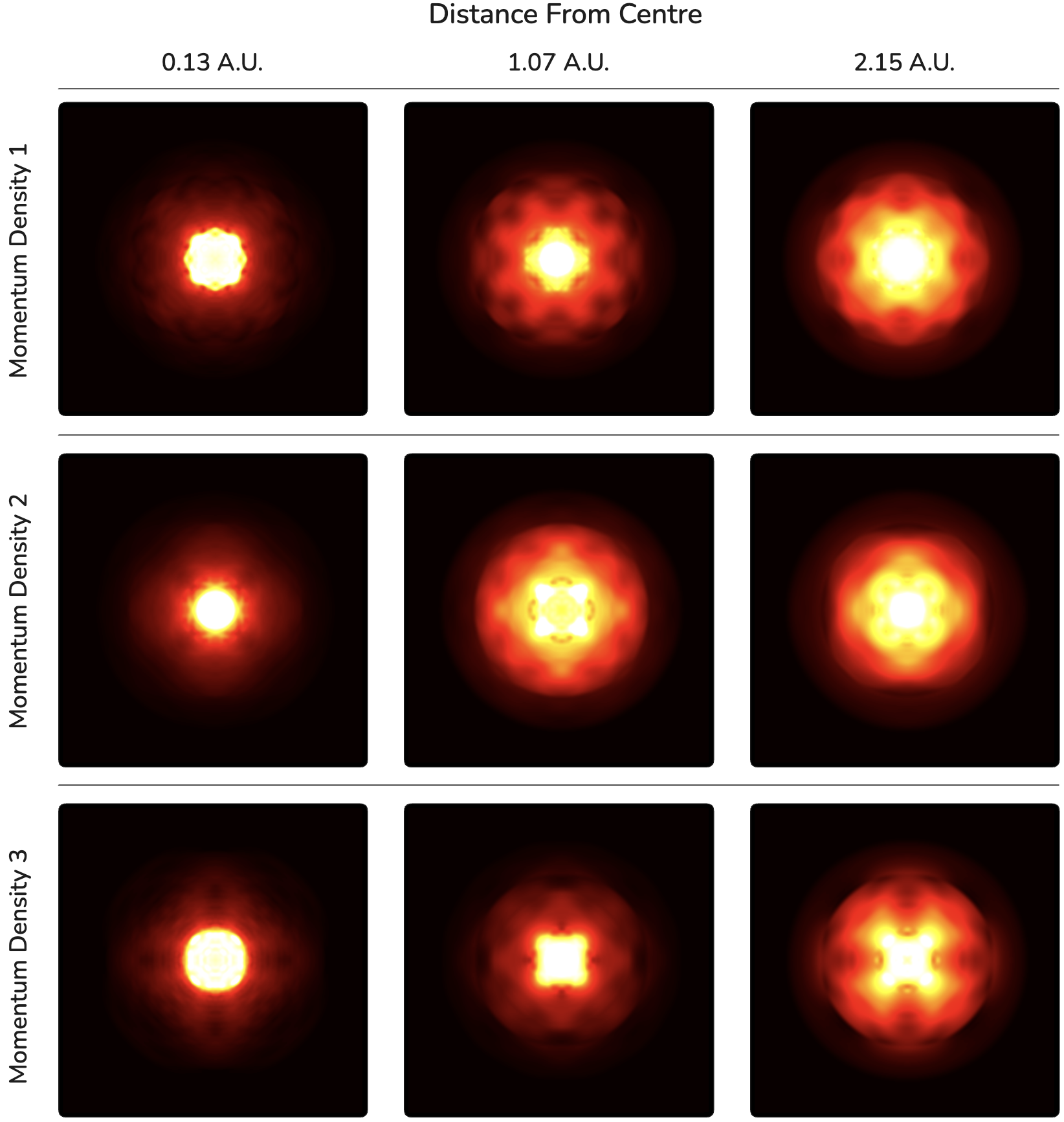}
  \caption{\footnotesize \textit{Comparing synthetic TPMD slices continuously evolved via Dynamic Mode Decomposition.}}
  \label{AppendixFig:SynthTPMDsComparison}
\end{figure}

\newpage
\section{Simulated Experimental Conditions Validation \label{Appendix:SimExpVal}}

\begin{figure}[htp!]
  \centering  
  \includegraphics[width=0.98\textwidth]{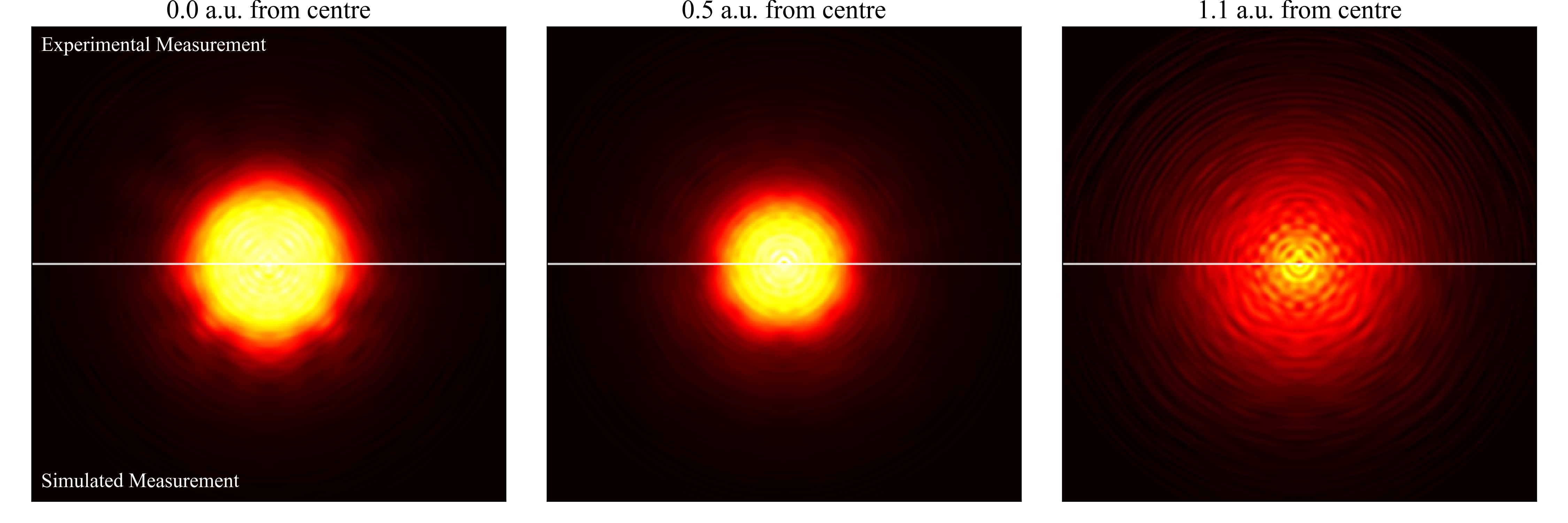}
  \includegraphics[width=0.98\textwidth]{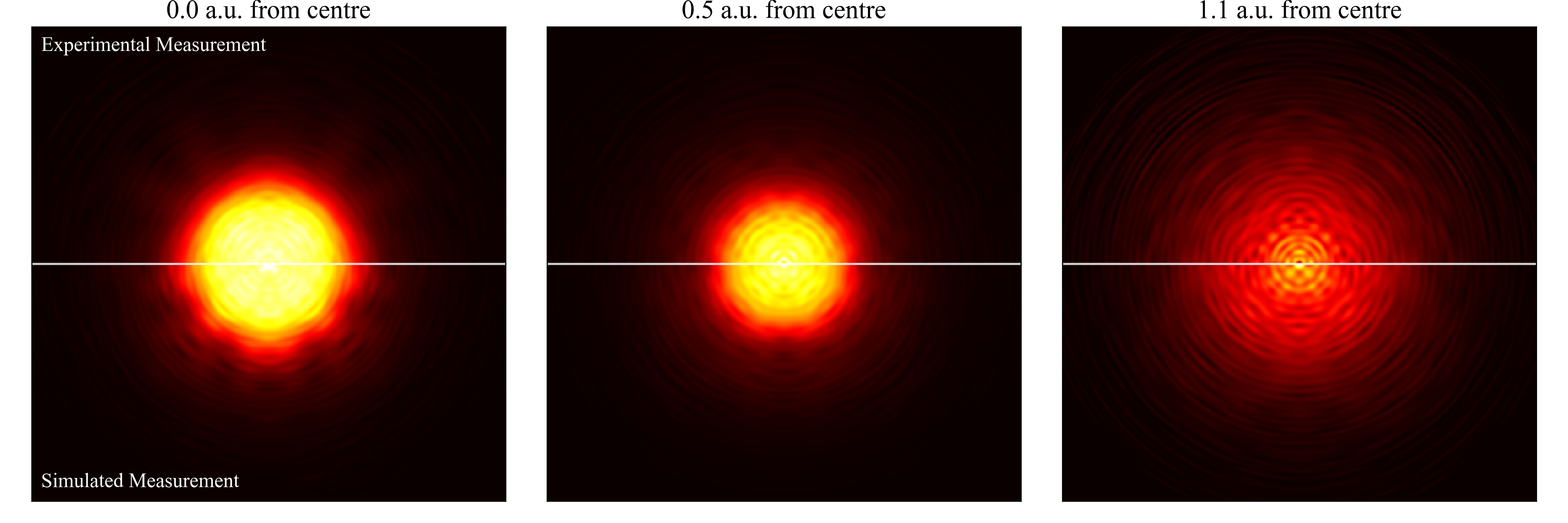}
  \includegraphics[width=0.98\textwidth]{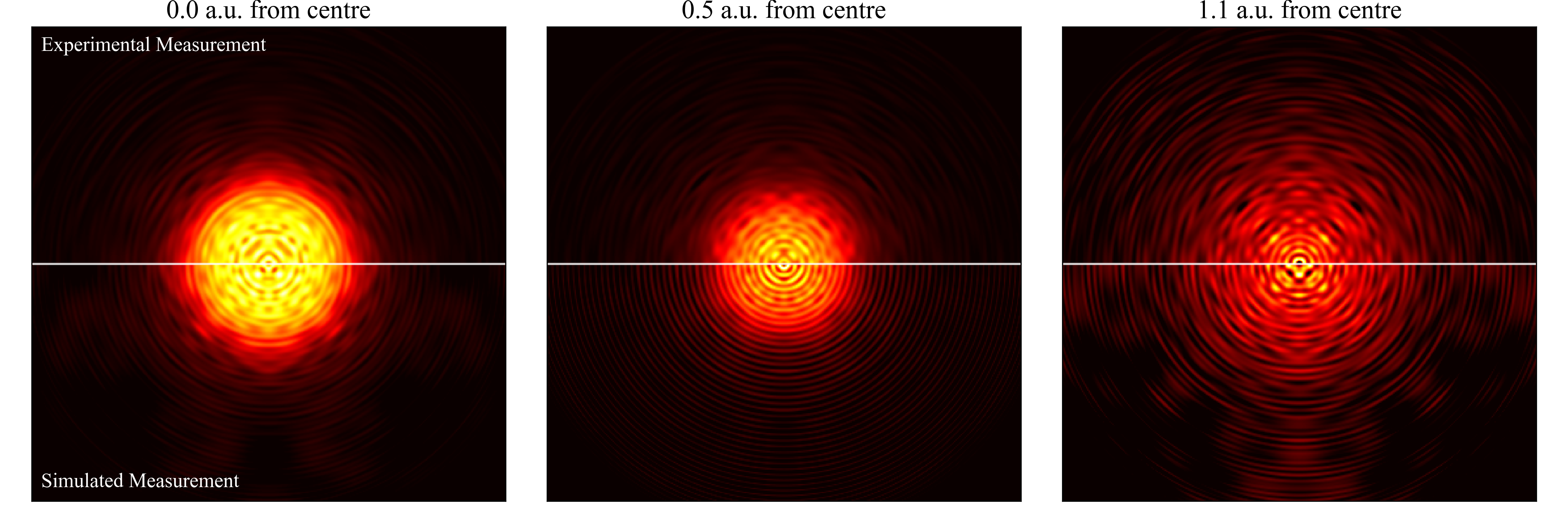}

  \caption{\footnotesize \textit{ZrZn$_2$ 3D TPMD reconstruction at varying counts (165M, 100M, 10M), comparing the experimental measurement with the simulated experimental measurement through the corresponding slice. The simulated experimental data used 60\% greater convolution and 40\% of the true counts (i.e. greater noise) to give comparable degradation quality to the actual experimental data.}}
  \label{AppendixFig:TPMDvsCounts}
\end{figure}

\end{document}